\documentclass[10pt,letterpaper]{article}
\usepackage[top=0.85in,left=2.75in,footskip=0.75in]{geometry}

\usepackage{amsmath,amssymb}

\usepackage{changepage}

\usepackage[utf8x]{inputenc}

\usepackage{textcomp,marvosym}

\usepackage{cite}

\usepackage{nameref,hyperref}

\usepackage{microtype}
\DisableLigatures[f]{encoding = *, family = * }

\usepackage[table, dvipsnames]{xcolor}

\usepackage{array}

\newcolumntype{+}{!{\vrule width 2pt}}

\newlength\savedwidth

\usepackage{setspace} 

\raggedright
\usepackage[aboveskip=1pt,labelfont=bf,labelsep=period,justification=raggedright,singlelinecheck=off]{caption}

\makeatletter
\renewcommand{\@biblabel}[1]{\quad#1.}
\makeatother

\usepackage{lastpage,fancyhdr,graphicx}
\usepackage{epstopdf}
\fancyheadoffset[L]{2.25in}
\fancyfootoffset[L]{2.25in}
\usepackage{bbm}
\newcommand{\trial}[1]{#1^{(r)}}
\newcommand{\mbf}[1]{\mathbf{#1}}
\newcommand{\bs}[1]{\boldsymbol{#1}}
\def\R{\mathbb{ R}}

\begin{document}
\vspace*{0.2in}

\begin{flushleft}
{\Large
\textbf\newline{Cross-population coupling of neural activity based on Gaussian process current source densities} 
}
\\
Natalie Klein\textsuperscript{1, 2\textcurrency},
Joshua H. Siegle\textsuperscript{3},
Tobias Teichert\textsuperscript{4},
Robert E. Kass\textsuperscript{1, 2, 5*}
\\
\bigskip
\textbf{1} Department of Statistics and Data Science, Carnegie Mellon University, Pittsburgh, Pennsylvania, USA
\\
\textbf{2} Machine Learning Department, Carnegie Mellon University, Pittsburgh, Pennsylvania, USA
\\
\textbf{3} MindScope Program, Allen Institute, Seattle, Washington, USA
\\
\textbf{4} Departments of Psychiatry and Bioengineering, University of Pittsburgh, Pittsburgh, Pennsylvania, USA
\\
\textbf{5} Neuroscience Institute, Carnegie Mellon University, Pittsburgh, Pennsylvania, USA

* kass@stat.cmu.edu

\textcurrency Current address: Statistical Sciences Group, Los Alamos National Laboratory, Los Alamos, New Mexico, USA

\end{flushleft}
\section*{Abstract}
Because local field potentials (LFPs) arise from multiple sources in different spatial locations, they do not easily reveal coordinated activity across neural populations on a trial-to-trial basis. As we show here, however, once disparate source signals are decoupled, their trial-to-trial fluctuations become more accessible, and cross-population correlations become more apparent. To decouple sources we introduce a general framework for estimation of current source densities (CSDs). In this framework, the set of LFPs result from noise being added to the transform of the CSD by a biophysical forward model, while the CSD is considered to be the sum of a zero-mean, stationary, spatiotemporal Gaussian process, having fast and slow components, and a mean function, which is the sum of multiple time-varying functions distributed across space, each varying across trials. We derived biophysical forward models relevant to the data we analyzed.
In simulation studies this approach improved identification of source signals compared to existing CSD estimation methods. Using data recorded from primate auditory cortex, we analyzed trial-to-trial fluctuations in both steady-state and task-evoked signals. We found cortical layer-specific phase coupling between two probes and showed that the same analysis applied directly to LFPs did not recover these patterns. We also found task-evoked CSDs to be correlated across probes, at specific cortical depths. Using data from Neuropixels probes in mouse visual areas, we again found evidence for depth-specific phase coupling of primary visual cortex and lateromedial area based on the CSDs.

\section*{Author summary}
To better understand information processing in the brain, it is important to identify situations in which neural activity is  coordinated across populations of neurons, including those in distinct layers of the cortex. Bulk population activity results in voltage changes across extracellular electrodes, but in raw form such voltage recordings can be hard to analyze and interpret. 
In this paper, we develop a novel framework for locating the sources of currents that produce the measured voltages, and decomposing those currents into interpretable components. We use this flexible framework to develop statistical methods of analysis, and we show that our methodology can be more effective than existing techniques for current source localization. We apply the method to extracellular recordings in two contexts: primate auditory cortex in response to tone stimuli, and mouse visual cortex in response to visual stimuli. In both cases, we get results that are useful for understanding cross-population activity while being difficult or impossible to obtain using the raw voltage signals. We thereby demonstrate the broad utility of our approach for identifying coordinated neural activity based on extracellular voltage recordings.


\section*{Introduction}
Local field potentials (LFPs) recorded from multiple electrodes are influenced in part by interactions among populations of neurons. However, because they involve a variety of post-synaptic potentials near the recording electrode~\cite{buzsaki2012origin,einevoll2013modelling}, and may also be affected by more distant sources~\cite{linden2011modeling,kajikawa2011local,herreras2016local,pesaran2018investigating}, using LFPs to identify cross-population co-activity is  difficult: it can be hard to disentangle the many signals mixed together in LFPs, which may have multiple timescales and likely arise from distributed current sources contaminated by noise. We have developed and investigated a statistical modeling approach that incorporates the biophysical relationship between cellular current flow and measured voltages, and assumes the current source densities are stochastic, driven by several distinct processes, each operating at a different timescale. The goal of our work is to identify trial-to-trial covariation among features of the current source densities, using such covariation as an indicator of coordinated activity across neural populations. As we show, relationships across populations can become apparent when analyzing current source densities even though they are invisible among LFPs.

Traditional current source density (CSD) estimation is based on a particular forward model, and it applies second derivatives along a discrete grid of locations~\cite{pitts1952investigations,nicholson1971field}. This not only restricts its application to LFPs from evenly-spaced recording electrodes, but it also makes traditional CSD estimation sensitive to noise because second derivatives of noisy functions are unstable~\cite{dennis1996numerical}. Our approach is closer, in spirit, to a pair of more recent suggestions, termed inverse CSD~\cite{pettersen2006current} and kernel CSD~\cite{potworowski2012kernel}, but includes important additional structure. We apply a forward model while assuming there are both steady (spontaneous) and transient (evoked) currents, with the steady currents having fast and slow components; these currents are then mapped to the voltage detected at each electrode using a forward model, and we assume noise is added to produce the LFP.
In schematic form, our modeling approach can be written,
\begin{eqnarray*}
\mbox{CSD } &=& \mbox{ transient } + \mbox{ slow } + \mbox { fast }\\
\mbox{LFP } &=& \mbox{ forward(CSD) } + \mbox{ noise}
\end{eqnarray*}
where the slow and fast terms on the right of the CSD equation are spatiotemporal Gaussian processes 
with different timescales.  From this intuitive conception
we have constructed a framework for finding trial-to-trial correlation, across populations, both in transient current increases and in steady but rapidly changing currents (oscillations). This framework, spelled out in Eqs~(\ref{eq:phi})-(\ref{eq:lfpgenmodel}) in the next section, is able to improve identification of currents substantially while also offering flexibility: it can be used to analyze data from electrodes spaced unevenly in 1, 2, or 3 spatial dimensions, and it allows parameter-tuning in conjunction with model fitting. Furthermore, even though the non-evoked activity is assumed to have only two components, the model is able to reproduce the spectra of measured LFPs. Because the framework is built around Gaussian process current source densities, we use the acronym GPCSD to refer to the methodology and resulting CSD estimates.

The forward model maps the source space into a lower-dimensional observation space. Estimation of sources from data therefore creates an ill-posed inverse problem. GPCSD effectively regularizes by using a flexible Gaussian process structure with a relatively low-dimensional parameter space. We confirm here, using simulated data, that our approach can outperform  existing methods in recovering ground truth  CSDs, even when the modeling assumptions are incorrect.
We then illustrate potential findings by analyzing two data sets: LFPs from a pair of laminar probes in primate auditory cortex, in response to auditory cues, and LFPs from Neuropixels~\cite{jun2017fully} in distinct visual areas of the mouse, in response to visual stimuli~\cite{siegle2021survey}. In the case of the auditory data, we extracted dense spatial profiles of CSD activity operating at different time scales, and identified 10Hz phase coupling across populations at similar cortical depths, and also across different depths; these relationships were not recovered by applying phase coupling methods directly to the recorded LFPs. 
Previous work suggests that current sources are useful for understanding trial-to-trial variation~\cite{szymanski2011laminar}, so we also used CSDs to identify trial-to-trial correlation in evoked responses to an auditory stimulus~\cite{arieli1996dynamics}.
Using the LFPs from Neuropixels we found depth-specific phase coupling of primary visual cortex (V1) and lateromedial area (LM) in both theta and beta bands.
Together, the simulation and real data results demonstrate the possibility of inferring cooperative population activity from LFPs.

In the following sections, we first describe the general statistical formulation of our approach, including details about our forward modeling. This framework is general. To apply it in a given problem, choices must be made. We specify these particulars in the context of both simulated and real data. We then present our data-analytic results.

\section*{Results}

\subsection*{Statistical formulation of CSD estimation}
In this section, we describe our conception of the CSD as a spatiotemporal stochastic process, we outline the general form of the forward model, and we show how we adapt it to our one-dimensional and two-dimensional data applications.

\subsubsection*{CSD as a spatiotemporal process}
Biophysical forward models describe the transformation from currents to recorded voltages at any instant in time.
In this work, we use the term ``biophysical forward model'' to refer to a simplified forward model based on simplifying assumptions; in particular, we do not include details such as properties of specific neurons in the forward models.
Current flow across a cell membrane creates a current source or sink, and the resulting three-dimensional field potential can be derived using volume conductor theory. 
Because it is not possible to determine all the contributions of individual transmembrane currents from measured LFPs, the CSD at time $t$ and spatial point $s$ may be conceptualized as the average transmembrane current within a small area around $s$~\cite{einevoll2013modelling}. We let $c(s,t)$ denote the CSD, taking it to be a continuous function, which we will estimate from the LFP data. The biophysical model maps the CSD $c$ to the field potential, which we write as $\phi$, through a spatial linear operator $\mathcal{A}_s$, so that $\phi=\mathcal{A}_s[c]$. We use the subscript $s$ to emphasize the spatial operation. The operator
takes the form of an integral, so that the field potential at $(s,t)$ is given by
\begin{equation}
\phi(s,t)=\mathcal{A}_s[c](s, t) = \int a(s,s')c(s',t)ds'\label{eq:phi}
\end{equation}
for a suitable function $a(s,s')$ (the form of which will become clear).
We write the measured LFP as $\tilde{\phi}$ and assume it is equal to $\phi$ plus independent noise:
\begin{equation}
\tilde{\phi}(s,t)=\phi(s,t) + \epsilon(s,t). \label{eq:tildephi}
\end{equation}
Across repeated trials, in response to the same stimulus or producing the same behavior, the CSD will vary~\cite{arieli1996dynamics}. We use a superscript $r$ to identify relevant quantities on trial $r$, for $r=1,\ldots,N_r$, and we assume the CSD functions $\trial{c}(s,t)$ are independent realizations of a stochastic process. 
Finally, we assume the CSD $c(s',t)$ is the sum of three components operating at distinct timescales, an evoked transient mean $\mu(s',t)$, a slowly varying stationary process $\eta_1(s',t)$, and a more rapidly varying stationary process $\eta_2(s',t)$. This last process $\eta_2$ could be oscillatory.  Thus, on trial $r$ we have
\begin{align}
    \trial{c}(s', t) = \trial{\mu}(s', t) + \trial{\eta}_1(s', t) + \trial{\eta}_2(s', t), \label{eq:csdgenmodel}
\end{align}
so that $\trial{c}(s',t)$ becomes a realization of a stochastic process, the realizations being independent across trials. Because $\mathcal{A}_s$ is linear, on trial $r$ we get
\begin{align}
    \trial{\tilde{\phi}}(s, t) = \mathcal{A}_s[\trial{\mu}](s, t) + \mathcal{A}_s[\trial{\eta_1}](s, t) + \mathcal{A}_s[\trial{\eta_2}](s, t) + \trial{\epsilon}(s, t). \label{eq:lfpgenmodel}
\end{align}

Eqs~(\ref{eq:phi})-(\ref{eq:lfpgenmodel}), together with estimation of the current source density from the LFP data, define the fundamental components of the framework we have developed. 
By taking $\eta_1$ and $\eta_2$ to be Gaussian processes, the functions $\trial{c}(s',t)$ in Eq~(\ref{eq:csdgenmodel}) become Gaussian process current source densities, and we are able to infer features of them, on a trial-by-trial basis. 
More specifically, the free parameters in Eq~(\ref{eq:csdgenmodel}), together with a free parameter in our forward model, make up a parameter vector $\theta$ for a log likelihood function (see Eq~(\ref{eq:marglik})), and the current source density on trial $r$ is estimated using the collection of $\trial{\tilde{\phi}}(s,t)$ observations across the relevant grid of $(s,t)$ values (Eq~(\ref{eq:gppred})).
As we will show, this model can generate multiple time series similar to those observed in LFPs~\cite{buzsaki2012origin}.
The specifications of the forward model in Eq~(\ref{eq:phi}) and the three terms on the right of Eq~(\ref{eq:csdgenmodel}) are application-dependent. We discuss them below in the context of our two data sets.

\subsubsection*{Forward model details}
We first give an overview of a commonly-used three-dimensional forward model, which is relevant when three-dimensional LFP recordings are available. In the following section we describe additional \textit{a priori} physical models that are required to adapt the generative model and forward model to LFP measurements from one-dimensional and two-dimensional electrode arrays. 

Previous work on CSD estimation has established that an assumption of an isotropic, homogeneous medium with scalar conductivity $\varsigma$ is a reasonable approximation for cortical signals in primates~\cite{logothetis2007vivo}; see also the discussion of forward modeling assumptions in~\cite[Section 4.2.1]{pettersen2012extracellular}.
Using the quasi-static assumption, the relationship between the CSD $c$ and the LFP $\phi$ at a single time point (time index suppressed for clarity of notation) is governed by the Poisson equation~\cite{pitts1952investigations}:
\begin{align} \label{eq:poisson}
    \varsigma \nabla \cdot (\nabla \phi(x,y,z)) =
    \varsigma \left( \frac{\partial^2 \phi(x,y,z)}{\partial x^2} + \frac{\partial^2 \phi(x,y,z)}{\partial y^2} + \frac{\partial^2 \phi(x,y,z)}{\partial z^2} \right) = -c(x,y,z), 
\end{align}
where the spatial location $s$ is described in terms of coordinates $(x, y, z)$.
While this appears to give a formula for computing the CSD from the LFP, it requires detailed, accurate knowledge of the LFP in three dimensions, without which it fails to accurately recover the CSD~\cite{nicholson1975theory}.
In addition, estimated second derivatives are highly influenced by noise.
However, assuming an infinite volume conductor with negligible boundary conditions, the differential equation may be inverted to an integral equation which instead gives $\phi$ in terms of a linear integral operator $\mathcal{A}_s$ on $c$~\cite{nicholson1971field}:
\begin{align}
    \phi(x,y,z) = \mathcal{A}_s[c](s) \equiv - \frac{1}{4\pi\varsigma} \int \int \int \frac{c(x',y',z')}{\sqrt{(x-x')^2+(y-y')^2+(z-z')^2}} \, dx' \, dy' \, dz'. \label{eq:greens}
\end{align}

\subsubsection*{Adapting the model}
Though our generative model and forward model have thus far been discussed in terms of general processes over three dimensions, it is common for electrodes to be arrayed along only one or two dimensions, which leaves unmeasured the effects in other dimensions. 
One advantage of the formulation we have described is that it can be adapted to these situations by introducing \textit{a priori} assumptions about the CSD in the unmeasured directions.

One-dimensional recordings are typically taken along a linear probe inserted perpendicular to the surface of the cortex, so we will index the one-dimensional spatial locations by a single coordinate $z$ representing depth along the probe.
We use an \textit{a priori} physical model in which the CSD is assumed constant in the dimensions perpendicular to the linear probe on a cylinder of radius $R$ around the probe and zero elsewhere~\cite{pettersen2006current}; previous work has shown deviations from this shape do not have a large impact on the results~\cite{nicholson1973theoretical,potworowski2012kernel}.
Additionally, as linear probes are typically inserted to penetrate the thickness of the cortex, we assume the CSD is nonzero only on an interval $a \le z \le b$ representing the thickness of the cortex.
These assumptions lead to the following \textit{a priori} physical model that describes the variation of the CSD in the $z$ direction through a  function $g(z, t)$ that varies over time and a single spatial dimension:
\begin{align} \label{eq:cylmodel}
    c(x,y,z,t) = g(z,t) \mathbbm{1}(x^2 + y^2 \le R) \mathbbm{1}(a \le z \le b).
\end{align}
In Eq~(\ref{eq:cylmodel}), $\mathbbm{1}(\cdot)$ is an indicator function that evaluates to 1 if the argument is true and 0 otherwise, and we have used the convention that $x = y = 0$ corresponds to the probe location.
Under this \textit{a priori} physical model, as shown in~\nameref{S1Text} and previously derived in~\cite{pettersen2006current}, the three-dimensional forward model of Eq~(\ref{eq:greens}) reduces to
\begin{align} \label{eq:onedimfwd}
    \phi(z, t) = \mathcal{A}_z[g](z, t)  
    \equiv - \frac{R}{2\varsigma} \int_a^b g(z', t) \underbrace{\left[ \sqrt{\left(\frac{z-z'}{R}\right)^2+1} - \sqrt{\left(\frac{z-z'}{R}\right)^2}\right]}_{a(z, z'; R)} \, dz'.
\end{align}
Thus, under this \textit{a priori} physical model, the one-dimensional LFP is the result of applying a linear operator $\mathcal{A}_z$ to $g$, where the weighting function $a(z, z'; R)$ (see Eq~(\ref{eq:phi})) decreases with distance $z - z'$ but also depends on the cylinder radius $R$.
In this case, our prior model on the CSD specifies the space of functions $g(z, t)$ with one spatial and one temporal dimension.

Two-dimensional LFP measurements are commonly collected using microelectrode arrays such as a Utah array or Neuropixels probe~\cite{jun2017fully,steinmetz2018challenges}. 
In the Neuropixels data we will analyze, the two-dimensional spatial locations represent depth and width along the probe, so we index the spatial location $s$ by coordinates $y$ for width and $z$ for depth.
Unlike with Utah arrays, for Neuropixels, sources behind the face housing the electrode contacts are unlikely to make contributions to the measured LFPs~\cite{buccino2019does}, so we assume they don't. We also assume zero charge in a small area in front of the probe face, which avoids singularities in the forward model.
Therefore, the \textit{a priori} physical model will assume that the CSD is constant in front of the face of the probe in the unmeasured $x$ direction for some distance $R$ after a space of $\tau$; that is, using the convention that $x = 0$ is the face of the probe and positive $x$ are in front of the probe, we assume
\begin{align} \label{eq:2dmodel}
    c(x,y,z,t) = g(y, z, t) \mathbbm{1}(\tau \le x \le R+\tau) \mathbbm{1}(a_z \le z \le b_z)\mathbbm{1}(a_y \le y \le b_y).
\end{align}
We note that this forward model is similar to previously derived models for multielectrode arrays~\cite{lkeski2011inverse, ness2015modelling}.
In this case, the three-dimensional forward model of Eq~(\ref{eq:greens}) reduces to (see~\nameref{S1Text} for derivation)
\begin{align} \label{eq:twodimfwd}
    \phi(y, z, t) &= \mathcal{A}_{y,z}[g](y, z, t) \nonumber \\
    &\equiv - \frac{1}{4\pi\varsigma}  \int_{a_z}^{b_z} \int_{a_y}^{b_y} g(y', z', t) \underbrace{\log\left( \frac{R + \tau + \sqrt{(R+\tau)^2 + r_y^2+r_z^2}}{\tau + \sqrt{\tau^2 + r_y^2+r_z^2})}\right)}_{a(y, y', z, z'; R, \tau) \, \text{where} \, r_y = y-y', \, r_z = z-z'} \,  dy' \, dz'.
\end{align}
The prior in this case specifies the space of functions $g(y, z, t)$ with two spatial dimensions and one temporal dimension.
We note that the form of the forward model is similar to existing two-dimensional forward models for slightly different measuring devices such as Utah arrays~\cite{potworowski2012kernel}, but with some differences due to the \textit{a priori} physical model we have assumed for the Neuropixels probe.

\subsection*{Modeling and implementation details}
The framework in Eqs~(\ref{eq:phi})-(\ref{eq:lfpgenmodel}) is very flexible and must be adapted to the particulars of each situation. In this section, we discuss specific distributional modeling choices for the two data sets we analyze here. We also provide implementation details for model estimation and inference. The Python code implementing the methodology presented in this paper is available at Zenodo~\cite{klein_code} and as a Python package on pyPI \url{https://pypi.org/project/gpcsd/}, with full source code at \url{https://github.com/natalieklein/gpcsd}.

\subsubsection*{CSD prior model}
In Eq~(\ref{eq:lfpgenmodel}), the noise $\epsilon$ is modeled as zero-mean Gaussian, independent over space and time, with variance $\sigma^2$ (though extensions to spatially or temporally varying noise variances are easily included without complicating the computational framework).
Details of the transient mean $\trial{\mu}=\trial{\mu}(s',t)$  differ depending on the application, so specific choices will be discussed in context; in the remainder of this section, we will assume $\trial{\mu}$ is known and that it absorbs the non-stationary activity. The $\eta$ processes are then modeled as independent zero-mean Gaussian processes that are stationary in time and space, with realizations being independent across trials. To complete the model specification, we assume the spatiotemporal covariance for the sum of the slow and fast processes is 
separable in space and time.
That is, the covariance at $(s, t)$ and $(s', t')$ decomposes as
\begin{align}
    k(s,t; s',t') = k^s(s,s') [k_{(1)}^t(t,t') + k_{(2)}^t(t,t')].
\end{align}
Compared with a non-separable covariance structure,  separability greatly reduces the number of free parameters in the model and yields large savings in computation time. As we show in our simulation and real-data results, it remains possible to fit spatially distinct sources that have different profiles in time.
While in principle, processes at multiple time scales may not be identifiable, we distinguish between the slow and fast processes using priors on the lengthscale parameters.

One of the advantages of our generative model framework is that various spatial and temporal covariance functions can be used to include different prior beliefs about the underlying process, including its smoothness in space and time or periodicity; an overview of commonly-used covariance functions is given in~\cite[Ch. 4]{rasmussen2006gaussian}.
In our simulations and data analysis, we use the following covariance function structure.
The spatial covariance function is a unit variance squared exponential, or SE, with one lengthscale for each of the $D$ spatial dimensions:
\begin{align*}
    k^s(s,s') = \exp\left( - \sum_{d=1}^D \frac{(s_d - s'_d)^2}{2 \ell_{s,d}^2}\right).
\end{align*}
The SE covariance function reflects the prior belief that the CSD is smooth over space with the lengthscale $\ell_{s,d}$ governing the frequency of variation along spatial dimension $d$.
The slow-timescale temporal covariance function is also SE, 
\begin{align*}
    k_{(2)}^t(t,t') = \sigma^2_2 \exp\left( - \frac{(t - t')^2}{2 \ell_{t,2}^2}\right),
\end{align*}
with $\sigma^2_2$ representing the marginal variance of the slow-timescale process and $\ell_{t,2}$ representing the frequency of variation.
The fast-timescale exponential covariance function can capture rougher, faster variations,
\begin{align*}
    k_{(1)}^t(t,t') =  \sigma^2_1 \exp\left( - \frac{|t - t'|}{\ell_{t,1}}\right), 
\end{align*}
and has marginal variance $\sigma^2_1$ and lengthscale $\ell_{t,1}$.
Note that the spatial covariance function has unit variance to avoid identifiability issues with the marginal temporal variances.

Denoting by $\mbf{K}_\mbf{s,t;s',t'}$ the spatiotemporal covariance matrix of the combined fast and slow CSD processes evaluated for vectors of spatial and temporal locations $\mbf{s}$, $\mbf{s'}$, $\mbf{t}$, and $\mbf{t'}$, the covariance matrix has the following structure:
\begin{align*}
    \mbf{K}_\mbf{s,t;s',t'} = \mbf{K}^s_{\mbf{s,s'}} \otimes \left(\mbf{K}^t_{(1) \mbf{t,t'}} + \mbf{K}^t_{(2)  \mbf{t,t'}}\right).
\end{align*}
Exploiting the covariance matrix structure allows faster matrix inversion that would be possible on a general spatiotemporal covariance matrix~\cite{saatcci2012scalable}.

\subsubsection*{Joint distribution of CSD and LFP}
We have now specified the generative model framework for the CSD in addition to forward models relating the CSD to the LFP for one-, two-, and three-dimensional recordings.
Before discussing how the CSD can be predicted using this model or how to fit the model, we give an overview of what the generative model and forward model imply about the joint distribution of the latent CSD and the observed LFPs.

Assuming we are working with LFPs measured in one or two spatial dimensions, our generative model provides the form of $g(s, t)$ describing the spatial and temporal variation of the CSD in the measured dimensions.
The generative CSD model gives $g$ as a Gaussian process with mean function $\mu$ and covariance function $k^s [k_{(1)}^t + k_{(2)}^t]$, which we denote as
\begin{align*}
    g \sim GP(\mu, \, k^s [k_{(1)}^t + k_{(2)}^t]).
\end{align*}
Much as in the finite-dimensional case, the application of a linear operator to a Gaussian process results in another process that is jointly Gaussian with the original process. We write the bilinear operator corresponding to $\mathcal{A}_s$, when applied to $k^s$, as $\mathcal{A}_s [k^s]\mathcal{A}_s^T$, so that
the marginal noiseless LFP process is given by
\begin{align*}
    \phi \sim GP(\mathcal{A}_s[\mu], \, \mathcal{A}_s [k^s]\mathcal{A}_s^T [k_{(1)}^t + k_{(2)}^t]).
\end{align*}
See Eq~(S.7) in~\nameref{S1Text}.

The joint Gaussian process governing the CSD and LFP processes implies that if we have noisy observations of $\tilde{\phi}$ at a discrete set of spatial locations $\mbf{s}$ and temporal locations $\mbf{t}$, then the joint distribution of the vectorized observations, denoted $\bs{\tilde{\phi}}_\mbf{s, t}$, and the vectorized CSD process observed at an arbitrary set of locations $\mbf{s'}$ and $\mbf{t'}$, denoted  $\mbf{g}_\mbf{s', t'}$, is multivariate Gaussian:
\begin{align}
    \label{eq:jointgp}
        \begin{bmatrix} 
            \mathbf{g}_{\mbf{s', t'}} \\ 
            \bs{\tilde{\phi}}_{\mbf{s, t}} 
        \end{bmatrix}
    \sim \mathcal{N} \left( 
        \begin{bmatrix} 
            \bs{\mu}_{\mbf{s', t'}} \\ 
            \mathcal{A}_s \bs{\mu}_{\mbf{s, t}} 
        \end{bmatrix}, \,
        \begin{bmatrix}
            \mbf{K}^s_{\mbf{s',s'}} \otimes \mbf{K}^t_{\mbf{t', t'}} &  \mathbf{K}^s_{\mbf{s',s}} \mathcal{A}_s^T \otimes \mbf{K}^t_{\mbf{t', t}}\\
            \mathcal{A}_s \mbf{K}^s_{\mbf{s,s'}} \otimes \mbf{K}^t_{\mbf{t, t'}} & \mathcal{A}_s \mathbf{K}^s_{\mbf{s,s}} \mathcal{A}_s^T \otimes \mbf{K}^t_{\mbf{t, t}} + \sigma^2 I 
    \end{bmatrix}
    \right),
\end{align}
where $\mbf{K}^t_{\mbf{t', t'}} = \mbf{K}^t_{(1),\mbf{t', t'}} + \mbf{K}^t_{(2),\mbf{t', t'}}$.
Note that Eq~(\ref{eq:jointgp}) refers to the joint distribution of CSD and LFP processes on a single trial, for notational clarity; to complete the model across trials, we assume the trials are independently and identically distributed.
The linear operator notation $\mathcal{A}_s \mathbf{K}^s_{\mbf{s,s}} \mathcal{A}_s^T$ indicates application of the linear operator jointly to both inputs of the covariance function used to compute $\mathbf{K}_{\mbf{s,s}}$, while $\mathcal{A}_s \mbf{K}^s_{\mbf{s,s'}}$ and $\mathbf{K}^s_{\mbf{s',s}} \mathcal{A}_s^T$ indicate application of the linear operator to only the first or second inputs, respectively.
Details of these calculations are given in~\nameref{S1Text}.
The joint distribution over discretely observed LFPs and the latent CSD process at any spatial and temporal coordinates of interest is used for tuning model parameters and for prediction of the CSD.

\subsubsection*{Estimation of model parameters}
Using properties of multivariate Gaussian distributions, Eq~(\ref{eq:jointgp}) implies that the marginal density of the observations for trial $r$, defined as
\begin{align}
    p\left(\trial{\bs{\tilde{\phi}}}_\mbf{s, t}\right) = \int p\left(  \trial{\bs{\tilde{\phi}}}_\mbf{s, t} | \mbf{g}_\mbf{s', t'}\right) p(\mbf{g}_\mbf{s', t'}) \, d \mbf{g}_\mbf{s', t'},
\end{align}
 is available in closed form:
\begin{align} \label{eq:margdist}
    \trial{\bs{\tilde{\phi}}}_\mbf{s, t} \sim \mathcal{N}(\mathcal{A}_s \bs{\mu}_\mbf{s, t}, \, \mathcal{A}_s \mathbf{K}^s_\mbf{s} \mathcal{A}_s^T \otimes \mathbf{K}^t_\mbf{t} + \sigma^2 \mbf{I}).
\end{align}
Taking all $N_r$ trials to be independent and identically distributed, we obtain the following log marginal likelihood function for the model parameters $\bs{\theta} = \left[ R, \left\{ \ell_{s,d} \right\}_{d=1}^D, \ell_{t,1}, \ell_{t,2}, \sigma_1^2, \sigma_2^2, \sigma^2 \right]$:
\begin{align} \label{eq:marglik}
    \log \mathcal{L}(\bs{\theta}) = - \frac{N_r}{2} \log \left( \left| \bs{\Sigma} \right| \right) - \frac{1}{2} \sum_{r=1}^{N_r} \left(\trial{\bs{\tilde{\phi}}}_\mbf{s, t}\right)^T \bs{\Sigma}^{-1} \trial{\bs{\tilde{\phi}}}_\mbf{s, t},
\end{align}
where $\bs{\Sigma} \equiv \mbf{K}^s_\mbf{s, s} \otimes \mbf{K}^t_\mbf{t, t} + \sigma^2 \mbf{I}$ depends implicitly on the covariance hyperparameters $[\left\{ \ell_{s,d} \right\}_{d=1}^D, \ell_{t,1}, \ell_{t,2}, \sigma_1^2, \sigma_2^2]$ and on the forward model hyperparameter $R$.
Note that direct inversion of $\bs{\Sigma}$ is costly and potentially unstable, but can be improved by exploiting the Kronecker product structure~(\nameref{S1Text}).

To estimate the model parameters, we use the log marginal posterior, which combines the log marginal likelihood with prior information about the model parameters:
\begin{align} \label{eq:gphyppost}
    p(\bs{\theta} | \bs{\tilde{\phi}}_{\mbf{s}}) \propto p(\bs{\tilde{\phi}}_\mbf{s}) p(\bs{\theta}).
\end{align}
While a fully Bayesian approach such as MCMC could be used to approximate the full posterior, we instead use maximum \textit{a posteriori} (MAP) estimation by maximizing the logarithm of Eq~(\ref{eq:gphyppost}) with respect to $\bs{\theta}$; in this case, the use of priors can be seen as a form of regularization on the log marginal likelihood that discourages unrealistic or unidentifiable parameter values.
In our implementation, we use inverse Gamma priors for the lengthscales in which the prior hyperparameters are chosen so that the 1\% and 99\% quantiles will fall at specific values, typically corresponding to the minimum and maximum observed distances in space or time~\cite{betanGP}.
We also use a similar inverse Gamma prior for the forward model parameter $R$, and use broad half-Normal priors for the marginal variances and the noise variance~\cite{betanGP}.

\subsubsection*{Predicting the CSD}
Given a fixed mean function and fixed values for model parameters for the covariance functions and forward model, predictions conditional on the observed values of $\tilde{\phi}$ can be made for any $\mbf{s', t'}$ for either $\phi$ or $g$; typically, we will be mostly interested in predicting the latent CSD described by $g$.
Using properties of multivariate Gaussians, Eq~(\ref{eq:jointgp}) yields the following distribution for the CSD conditioned on the observed LFPs:
\begin{align*}
    \mbf{g}_\mbf{s', t'} \left| \bs{\tilde{\phi}}_\mbf{s, t} \sim \mathcal{N}(\bs{\mu^*}, \mbf{K^*}) \right. .
\end{align*}
The conditional mean, which we use to predict $\mbf{g}_\mbf{s', t'}$ given $\bs{\tilde{\phi}}_\mbf{s, t}$, is 
\begin{align} \label{eq:gppred}
    \bs{\mu^*} &=  \bs{\mu}_\mbf{s', t'} + \left(\mathcal{A}_s \mathbf{K}^s_{\mbf{s',s}} \otimes \mathbf{K}^t_{\mbf{t',t}} \right) [\mathcal{A}_s \mathbf{K}^s_{\mbf{s,s}} \mathcal{A}_s^T \otimes \mathbf{K}^t_{\mbf{t,t}} + \sigma^2 \mbf{I}]^{-1} (\bs{\tilde{\phi}}_{\mbf{s, t}} - \mathcal{A}_s \bs{\mu}_\mbf{s, t}).
\end{align}
We will use the acronym GPCSD to refer to the process of using the Gaussian process conditional mean to predict the CSD at some set of spatial locations after tuning model parameters.

The structure of the covariance function also permits separate prediction of the fast- and slow-timescale components.
Let $\mbf{a} = [\mathcal{A}_s \mathbf{K}^s_{\mbf{s,s}} \mathcal{A}_s^T \otimes \mathbf{K}^t_{\mbf{t,t}} + \sigma^2 \mbf{I}]^{-1} (\bs{\tilde{\phi}}_{\mbf{s, t}} - \mathcal{A}_s \bs{\mu}_\mbf{s, t})$.
Then the conditional mean decomposes as
\begin{align*}
    \bs{\mu^*} &= \bs{\mu}_\mbf{s', t'} + \left(\mathcal{A}_s \mathbf{K}^s_{\mbf{s',s}} \otimes [\mathbf{K}^t_{(1) \mbf{t',t}} + \mathbf{K}^t_{(2) \mbf{t',t}} ] \right) \mbf{a} \\
    &= \bs{\mu}_\mbf{s', t'} + \left(\mathcal{A}_s \mathbf{K}^s_{\mbf{s',s}} \otimes \mathbf{K}^t_{(1) \mbf{t',t}} \right) \mbf{a} + \left(\mathcal{A}_s \mathbf{K}^s_{\mbf{s',s}} \otimes \mathbf{K}^t_{(2) \mbf{t',t}} \right) \mbf{a},
\end{align*}
where the second term corresponds to the fast-timescale prediction and the third term corresponds to the slow-timescale prediction.

\subsection*{Performance assessment}

We used simulated data both to validate the GPCSD method and to compare it against two existing CSD methods: traditional CSD (tCSD) and kernel CSD (kCSD). 
The general idea for the simulations was to generate ground-truth CSDs, then apply the forward model to obtain LFPs; as in real data, the LFPs were observed at a discrete set of spatial locations.
The performance of each method in recovering the CSD from the generated LFPs was assessed by comparing the predicted CSD to the ground truth CSD.
First, we used a simple dipole-like CSD configuration to visualize the performance of GPCSD compared to previous CSD methods. Then, we simulated larger training and test sets from a Gaussian process model to quantify the relative performance of GPCSD to the other methods. Brief descriptions are given here, but full details of data generation, parameter estimates, and other settings are described in~\nameref{Methods}.

\subsubsection*{Performance on simple CSD dipole configuration}
The first set of simulation results demonstrates the ability of the GPCSD method to recover simple dipole-like CSD patterns even in the presence of noise, and shows qualitative differences between GPCSD, tCSD, and kCSD. 
We generated a simple CSD template comprised of two positive and two negative Gaussian-shaped bumps across one spatial and one temporal dimension, then passed it through the forward model to obtain a noiseless LFP profile; noisy versions of the LFPs were also generated by adding white noise.
The GPCSD method was applied with the mean function $\mu$ assumed to be zero and standard priors used for all estimated parameters, which included the forward model parameter $R$.
We note that only a single realization of the spatiotemporal LFP was used for estimation, demonstrating that GPCSD can be applied to single-trial data without repetitions.
Traditional CSD was applied directly to the observed LFPs at each time point, while kCSD was applied with the forward model parameter $R$ set to the ground-truth value and the other tuning parameters (basis width and noise variance) chosen by cross-validation over a two-dimensional grid of values using the Python toolbox kCSD for 1D CSD estimation~\cite{pettersen2006current,potworowski2012kernel}.
Because different CSD methods recover the CSD pattern up to some multiplicative constant, the true CSD and the CSD predictions were rescaled to have maximum absolute value equal to 1, similar to~\cite{chintaluri2019kcsd}.

\begin{figure}
    \centering
    \includegraphics[width=\linewidth]{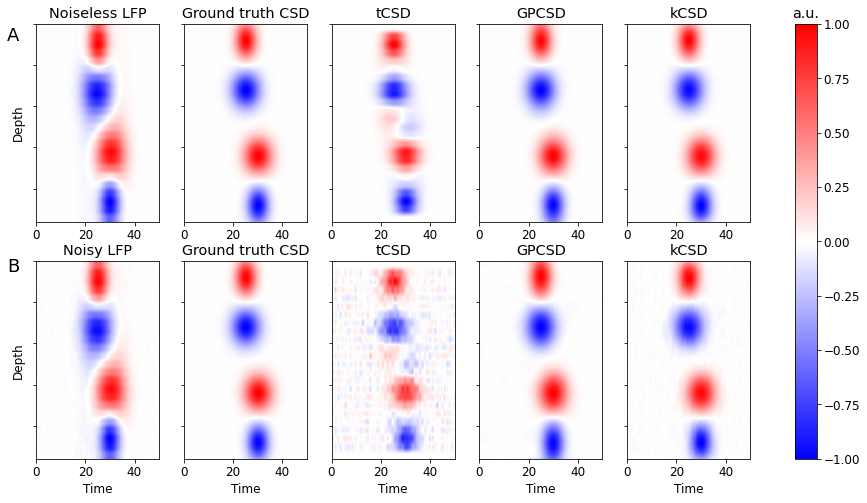}
    \caption{{\bf Comparing GPCSD to other methods on simulated CSD dipole.} (A) From left to right, ground-truth noiseless LFP, ground-truth CSD pattern, traditional CSD (tCSD) prediction, GPCSD prediction, and kernel CSD (kCSD) prediction. (B) Same as top row, but with noise added to the ground-truth LFP before CSD prediction. GPCSD accurately recovers the pattern compared to tCSD, and appears more robust to LFP noise than either tCSD or kCSD.} 
    \label{fig:CSDsimpletemplate}
\end{figure}

Figs \ref{fig:CSDsimpletemplate}A and \ref{fig:CSDsimpletemplate}B show the ground-truth CSD, noiseless LFP, and CSD predictions for tCSD, GPCSD, and kCSD, with the CSD predicted from the noiseless LFP in the top row (A) and from the noisy LFP in the bottom row (B).
In both cases, GPCSD reconstructed the ground truth accurately, though there were some small-amplitude artifacts due to the use of a stationary covariance function on a nonstationary ground-truth pattern.
Much more severe artifacts were present in the tCSD predictions, even in the noiseless case, and the performance of both tCSD and kCSD clearly degraded when white noise was added to the LFP while GPCSD recovered the pattern free of noise.

\subsubsection*{Quantifying performance on repeated trials}
\begin{figure}[ht!]
    \centering
    \includegraphics[width=0.8\linewidth]{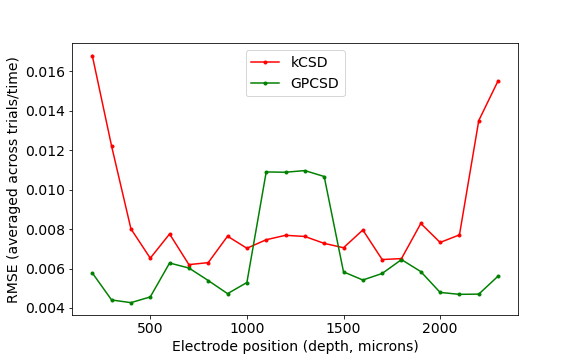}
    \caption{{\bf S1 Fig. Spatial distribution of errors in one-dimensional kCSD and GPCSD.} Average MSE (across trials/time) for the simulation study in the main text (with data generated from a GPCSD model). While our results suggested that kCSD and GPCSD overall perform similarly for this data, kCSD appears to have higher error at the edges while GPCSD has higher error in the center of the array.}
    \label{S1_Fig}
\end{figure}

To quantify the accuracy of GPCSD relative to the other methods, we generated multiple realizations of spatially one-dimensional CSDs from zero-mean spatiotemporal Gaussian process models, then passed these CSDs through the forward model to obtain LFPs.
The generated LFP trials were split into training and testing sets, each of size 50; the training set was used for selecting GPCSD and kCSD model parameters and the test set was used to evaluate the performance of each method in reconstructing the true CSD.
We again used the Python toolbox kCSD for 1D CSD estimation~\cite{pettersen2006current,potworowski2012kernel}.
Selected kCSD and GPCSD parameters are listed in~\nameref{Methods}. Because tCSD can only estimate the CSD at the interior electrode positions, we compared methods based on CSD predictions at these locations only.
For each test set trial, we computed the mean squared error (MSE) across all predicted space-time points for each trial. The MSEs averaged across trials were $7.38 \times 10^{-5}$, $4.64 \times 10^{-5}$, and 0.046 for GPCSD, kCSD, and tCSD, respectively. Paired $t$-tests indicated that GPCSD performed significantly better than tCSD ($p << 0.0001$) but there was weaker evidence that kCSD outperformed GPCSD ($p = 0.0003$). We note that if the GPCSD hyperparameters were fixed to their true values, the paired $t$-test for the difference in mean MSE between kCSD and GPCSD failed to detect a difference ($t=0.071, p=0.94$), suggesting that in this simulation, kCSD and GPCSD performed very similarly, but that results may depend on GPCSD hyperparameter optimization. In~\ref{S1_Fig}, we also show that the spatial distribution of error was different for kCSD and GPCSD, with kCSD tending to have higher errors at the edge relative to GPCSD and GPCSD tending to have higher errors in the center of the array than kCSD. Future work could explore this issue further, similar to~\cite{chintaluri2021we}. Additionally, we show in \nameref{S1Text} section ``Additional simulation results'' that GPCSD performed well in a similar simulation study but with the Gaussian process model mis-specified relative to the generating model.

\subsection*{Application to real data}
We analyzed LFP recordings from linear probes in primate auditory cortex and Neuropixels probes in the mouse visual system. In both sets of data we show how oscillatory activity within cortical layers can be coupled across populations of neurons. For these analyses we used phase locking value (PLV) to assess coupling of two phase angles and, following the development and methods in~\cite{torusgraphs}, what we call partial PLV to assess the coupling of two angles after conditioning on all other phase angles. As described in~\cite[Sections 4.1 and 4.4]{torusgraphs}, PLV is an angular analogue of correlation and partial PLV is an analogue of partial correlation. Torus graphs describe connectivity patterns across multiple angular random variables analogously to the way Gaussian graphical models describe connectivity among real-valued random variables. We also used the auditory data to illustrate the way GPCSDs can reveal coupling of transient activity across populations.

\subsubsection*{Application to auditory LFPs from laminar probes}
The auditory LFP data was collected from two simultaneously recording laminar probes inserted 3mm apart in primate primary auditory cortex (A1).
The first probe was located centrally in A1 and the second probe was located more medially and closer to the boundary of A1 with the medio-lateral belt, so we will refer to the probes as the lateral and medial probes, respectively. The medial probe had lower response threshold, shorter multi-unit activity (MUA) latencies, and overall stronger current sinks and sources than the lateral probe.
The recordings contained 2,509 trials in response to a collection of 11 pure tones with varying inter-stimulus intervals (ISIs) and fundamental frequencies on each trial; see~\nameref{Methods} for a full description of the data.
For each probe, we estimated putative cortical depth based on examination of the LFP and spiking activity. For interpretation of the results, each electrode contact was assigned to one of three depth levels (superficial, medium, or deep).
First, we investigated post-stimulus spectral power and phase coupling based on GPCSD predictions, assuming a zero-mean Gaussian process, from the steady LFP activity (activity during each trial but with the average evoked LFP response subtracted out prior to GPCSD modeling). 
Second, to study correlations in transient activity,
we estimated the CSD evoked response by first predicting the CSD during the trial using a fixed GPCSD model with hyperparameters estimated using the baseline pre-trial data, then taking the average across trials to obtain an evoked CSD pattern. 
We used spatially and temporally localized CSD components of the evoked response to detect correlated trial-to-trial variation in evoked response timing.

\subparagraph{Auditory steady activity}
To analyze steady activity, we first subtracted the LFP average evoked response (mean across trials) from each trial, then used the baseline period (100 ms before tone onset until tone onset) to estimate GPCSD model parameters for a zero-mean process separately for each probe.
We then predicted both the CSD and noiseless LFP at the original electrode positions and at time points from tone onset to 500ms after tone onset.

\begin{figure}[ht!]
    \centering
    \includegraphics[width=\linewidth]{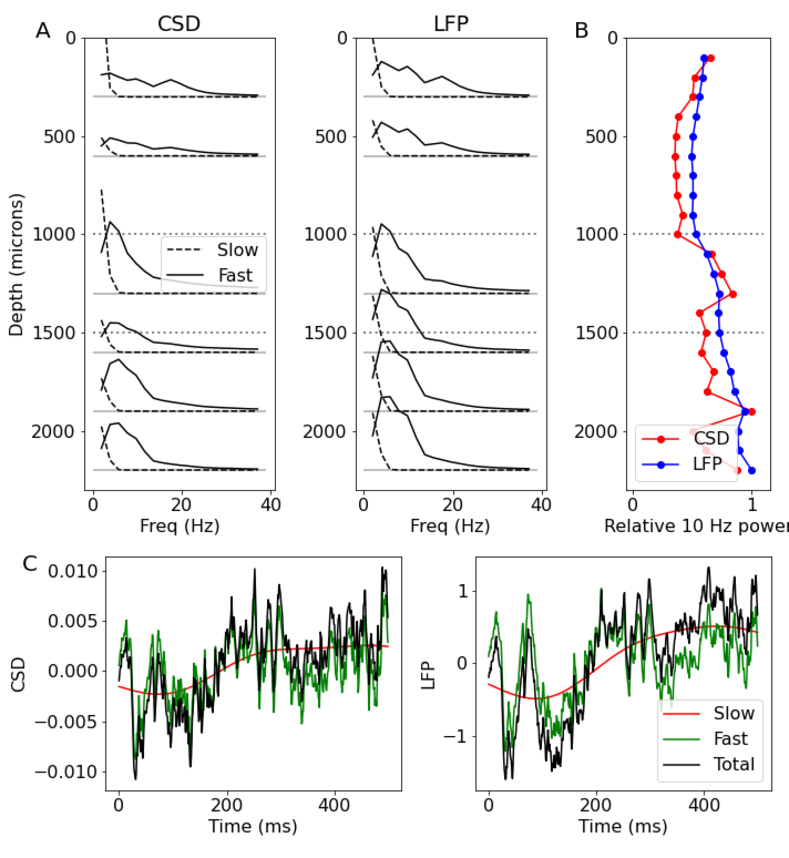}
    \caption{{\bf Spatial, spectral, and temporal properties of LFP and GPCSD.} (A) Periodograms (averaged across trials) showing power spectra at six cortical depths along the lateral probe for CSD (left) and LFP (right), computed for estimated fast-timescale processes (solid) and slow-timescale processes (dashed). Horizontal grey dotted lines indicate approximate boundaries between superficial, medium, and deep cortical layers. (B) Relative 10Hz power for CSD (red) and LFP (blue) as a function of cortical depth. (C) Time courses for a single trial for an electrode in the middle cortical layers, broken into slow, fast, and total, for CSD (left) and LFP (right). While the temporal components were similar in LFP and CSD, they did exhibit differences due to the different spatial properties of the CSD.}
    \label{fig:CSDspectra}
\end{figure}

We computed a periodogram for each trial separately for the fast and slow timescales, then averaged the resulting periodograms across trials.
Fig~\ref{fig:CSDspectra}A shows the trial-averaged periodograms for each timescale (solid: fast timescale, dashed: slow timescale) for both the CSD and LFP at six different depths along the lateral probe (results were similar for the medial probe).
Dotted horizontal lines indicate boundaries between superficial, medium, and deep layers.
For the fast timescale processes, the CSD and LFP periodograms were similar, but there were discernible differences, with the CSD periodograms exhibiting greater variation across depths. There were clear peaks around 10 Hz at several depths, with CSD seeming to be relatively stronger in the middle depth, around 1300 microns, and again in a deeper layer, around 1900 microns.  Fig~\ref{fig:CSDspectra}B displays power (relative to maximum) for the 10Hz frequency band for CSD (red) and LFP (blue). Again, while the CSD and LFP 10 Hz power plots remain similar, the distinctions across depths are more clearly visible in the CSD plot.
Fig~\ref{fig:CSDspectra}C shows example time courses for a single trial and a single electrode in the middle cortical layers for both CSD and LFP, decomposed into slow and fast components. While the temporal properties of the slow and fast components appear similar in both CSD and LFP, there are some differences in the values due to the spatial deconvolution of the CSD.

\begin{figure}[ht!]
    \centering
    \includegraphics[width=\linewidth]{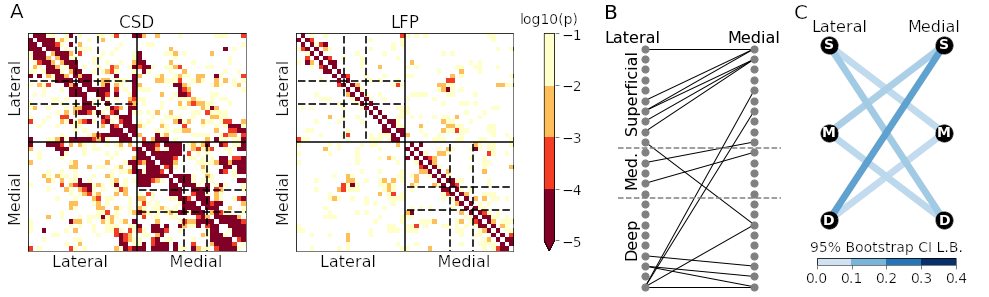}
    \caption{{\bf Phase coupling graphs in auditory CSD and LFP.} (A) Results of edgewise torus graph phase coupling inference both within- and across-probe for CSD (left) and LFP (right); depth along each probe is increasing left to right and top to bottom, and dashed lines indicate approximate boundaries between superficial, medium, and deep cortical layers.
    Colored entries correspond to edges, with color representing the $\log_{10}$ of the $p$-value. 
    Within-probe, the LFP had edges primarily along the diagonal, while the CSD contained more edges, including some connections across superficial, medium, and deep layers.
    Between probes, the CSD torus graph contained a noticeable set of edges, primarily along the diagonal, while the LFP torus graph had very few edges between probes.
    (B) Graph showing significant between-probe CSD torus graph edges, with lateral probe nodes ordered by depth in the left column and medial probe nodes ordered by depth in the right column; dashed lines indicate approximate boundaries between superficial, medium, and deep cortical layers.
    Many of the cross-probe connections occurred near the same depth on both probes, though there appear to be some edges connecting lateral probe deep layers to medial probe superficial layers.
    (C) Simplified graph between superficial (S), medium (M), and deep (D) cortical layers. Edge color corresponds to a 95\% bootstrap confidence interval lower bound for the partial PLV value (reflecting coupling strength, which falls between 0 and 1). The strongest cross-probe connection was between the deep layers of the lateral probe and the superficial layers of the medial probe.}
    \label{fig:CSDTGgraph}
\end{figure}

Next, we assessed phase coupling for oscillations centered at 10 Hz by using a bandpass filter centered at 10 Hz along with the Hilbert transform to extract instantaneous phases.
In both the CSD and LFP, the mean pairwise PLV across all electrode locations (both within-probes and between-probes) appeared to increase monotonically from before the stimulus until 100 ms after stimulus, then remained high and nearly constant until about 300 ms after stimulus, so we selected a time point during the later period (250 ms after stimulus) to investigate phase coupling.
We used torus graphs to construct a multivariate phase coupling graph describing connectivity among all 48 nodes (24 from each probe). 
The overall test of the null hypothesis of no edges between the two probes was significant for both the LFP and CSD phases ($p < 0.0001$).
Fig~\ref{fig:CSDTGgraph}A shows connectivity matrices for within- and between-probe connections, with edges colored by edgewise $\log_{10}$ $p$-value.
Horizontal dashed lines indicate the separation between superficial, medium, and deep cortical layers.
Within-probe, most connections tended to be the near the diagonal, particularly for the LFP, where there was strong evidence for edges mostly along the diagonal.
By inspection of the lower right quadrant of Fig~\ref{fig:CSDTGgraph}A, we also note that the medial probe CSD indicates a large number of strong connections from its deep to its superficial layers.
Across-probes, the torus graph based on LFP had very few edges, while CSD torus graph had many edges, with a noticeable spatial structure to the edge pattern.
Fig~\ref{fig:CSDTGgraph}B shows the between-probe connectivity matrix in graphical form, with edges shown for edgewise $p < 0.01$ (Bonferroni corrected for the number of edges tested).
Most connections between probes occurred between the same depths on each probe, but there was also very strong evidence for phase locking between the
lateral probe deep layers and the medial probe superficial layers.
While this result, taken in the context of deep-to-superficial edges within the medial probe, may suggest cross-layer connections between the medial and lateral areas, we note that the underdetermined nature of CSD estimation implies that the activity assigned to the deepest layers could potentially be coming from a deeper brain structure.
Fig~\ref{fig:CSDTGgraph}C shows a simplified version of the graph, taking only the center electrode in each depth range as the superficial (S), medium (M), and deep (D) nodes.
In this graph, the edges are colored by the lower bound of a 95\% bootstrap confidence interval on the partial PLV value, a statistic that falls between 0 and 1 and, much like PLV, quantifies the strength of the phase coupling.
The strongest edge connected lateral probe deep layers to medial probe superficial layers.

\subparagraph{Auditory transient activity}
In the steady activity analysis, we assumed that a single evoked response (estimated by the mean across trials) was shared across all trials and that any residual variation was not part of an evoked response. 
However, it is more likely that the transient evoked response varies from trial to trial, and this variation could indicate a different type of neural communication than the steady activity.
While previous work has demonstrated that trial-to-trial time lags in LFPs can be recovered on a coarse spatial scale based on steady state recordings~\cite{adhikari2010cross}, here we investigate trial-to-trial time lag variation in transient, nonstationary activity resolved to a fine spatial scale through the estimated CSD.
With the steady activity GPCSD model parameters fixed to the values estimated from the pre-stimulus baseline data, we first estimated a CSD evoked response function shared across all trials, then separated it into multiple CSD evoked components to investigate trial-to-trial timing and variation of specific evoked components localized in space and time.

\begin{figure}[ht!]
    \centering
    \includegraphics[width=\linewidth]{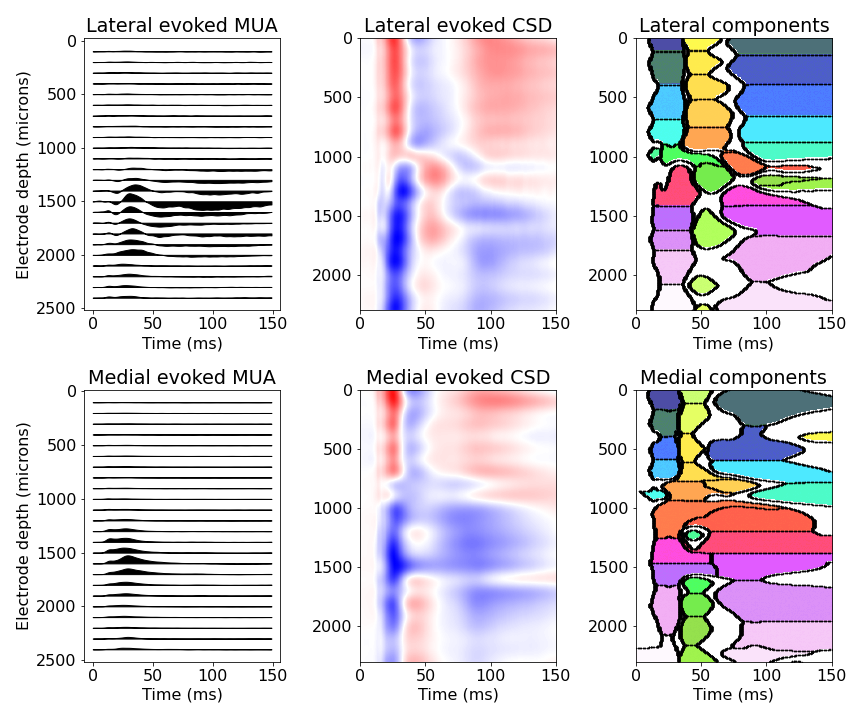}
    \caption{{\bf Estimated CSD evoked response components.} Left to right: Trial-averaged multi-unit activity (MUA) relative to baseline, estimated CSD evoked response, components returned by the image segmentation (colors correspond to arbitrary cluster number). Top row corresponds to the lateral probe and bottom row to the medial probe. The evoked responses for both probes have similar features but slightly different spatial and temporal properties.
    In particular, both the MUA and CSD evoked responses indicate that the evoked response begins earlier in the medial probe than the lateral probe.}
    \label{fig:CSDcomponents}
\end{figure}

To estimate the latent CSD mean function shared across all trials, we first estimated the CSD timecourses on each trial using the GPCSD model with parameters estimated from the baseline period, then took the mean across trials to obtain an evoked CSD. 
To assess trial-to-trial variation, we separated the fitted CSD mean function into multiple CSD components, non-overlapping in space and time, by applying image segmentation techniques (see~\nameref{Methods}).
The center of mass of the absolute value of each component was used as a marker of the peak time and spatial location of the component.
Fig~\ref{fig:CSDcomponents} shows the estimated evoked response function for each probe along with the components identified by the segmentation algorithm.
The evoked responses and components were similar in time and space between the two probes, but there were some differences in the spatial and temporal profiles.
It appears that the medial probe evoked responses started slightly earlier than the lateral probe responses, which is also consistent with the trial-averaged multi-unit spiking activity.
Both probes exhibited a current inversion near a depth of 1000 microns that appeared to persist even as the activity fluctuated between positive and negative current over time.

\begin{figure}[ht!]
    \centering
    \includegraphics[width=\linewidth]{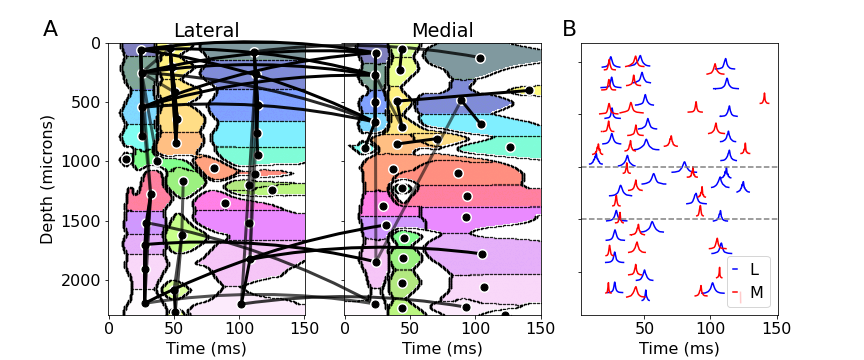}
    \caption{{\bf Per-trial shifts and correlations in CSD evoked response components.} (A) Spatiotemporal plot of evoked response components for the lateral probe (left) and the medial probe (right), colored to show separate components. Black circles represent centers of mass of each component, and an edge between two black circles indicates significant correlation in the per-trial shifts of the two components (darker, thicker edges indicate larger correlation values). Within each probe, the pattern of connections suggests that many components of the early evoked response (occurring before 60 ms) have related time shifts on a trial-to-trial basis; the lateral probe also has correlated time shifts in the later evoked components. 
    Between probes, there are connections between early evoked components at similar depths, with some evidence of shift correlations between lateral probe early components and medial probe later components.
    (B) Kernel density estimates of the peak times, across trials, of the evoked components in each probe with dashed lines marking putative cortical depth boundaries (separating superficial, medium, and deep layers). The medial probe responses generally precede the lateral probe responses across depths (confirmed by pairwise testing on the difference in means, $p < 0.001$ corrected).
    In addition, the ordering of responses across depths appears similar in each probe, with the earliest responses occurring in the superficial and deep layers, followed by the medium depth layers.}
    \label{fig:1dshiftgraph}
\end{figure}

Given the CSD evoked components, we then estimated a per-trial time shift for each component by maximizing the marginal likelihood of the data for each trial conditional on the estimated component shapes and estimated ongoing activity Gaussian process covariance function.
The estimated per-trial shifts for all components in both probes had across-trial means at 0 ms and across-trial standard deviations between 1ms and 3ms, depending on the component.
To determine components for which the shifts were related across trials, we computed correlations (across trials) between the point estimates of the shifts of each component from both of the probes.
Fig~\ref{fig:1dshiftgraph}A shows the components for the lateral and medial probes, with edges connecting the centers of mass of two components when there was significant correlation in the per-component shifts ($p < 0.001$, corrected; based on Fisher $z$-transform on the correlation coefficient).
Within-probe, it appears that many of the early evoked components (occurring before 60 ms) had correlated time shifts, indicating that time variation in early evoked components was related across trials.
Across probes, the component time shifts for most of the early evoked components were related to components near the same depth along the probe, though there were some connections from early lateral probe components to later medial probe components.
To quantify lagged relationships between probes in the earliest evoked responses (occurring before 40ms), we display kernel density estimates of the across-trial peak times for the responses in each probe in Fig~\ref{fig:1dshiftgraph}B along with dashed horizontal lines indicating depth boundaries (separating superficial, medium, and deep layers).
This figure indicates that across depth, medial probe evoked responses tended to precede lateral probe evoked responses; the differences in the mean peak times for evoked components at similar depths were also significantly different from zero ($p < 0.001$ corrected).
In addition, it appears that in both probes the superficial and deep layers had earlier peak times than the medium layers.
The earliest responses in A1 are typically observed in thalamo-recipient layers 4 and deep 3. It appears these responses are somewhat identified in the earliest responses occurring just above the dashed line separating superficial and medium layers.

\subsubsection*{Application to visual LFPs from Neuropixels probes}
\begin{figure}[ht!]
    \centering
    \includegraphics[width=\linewidth]{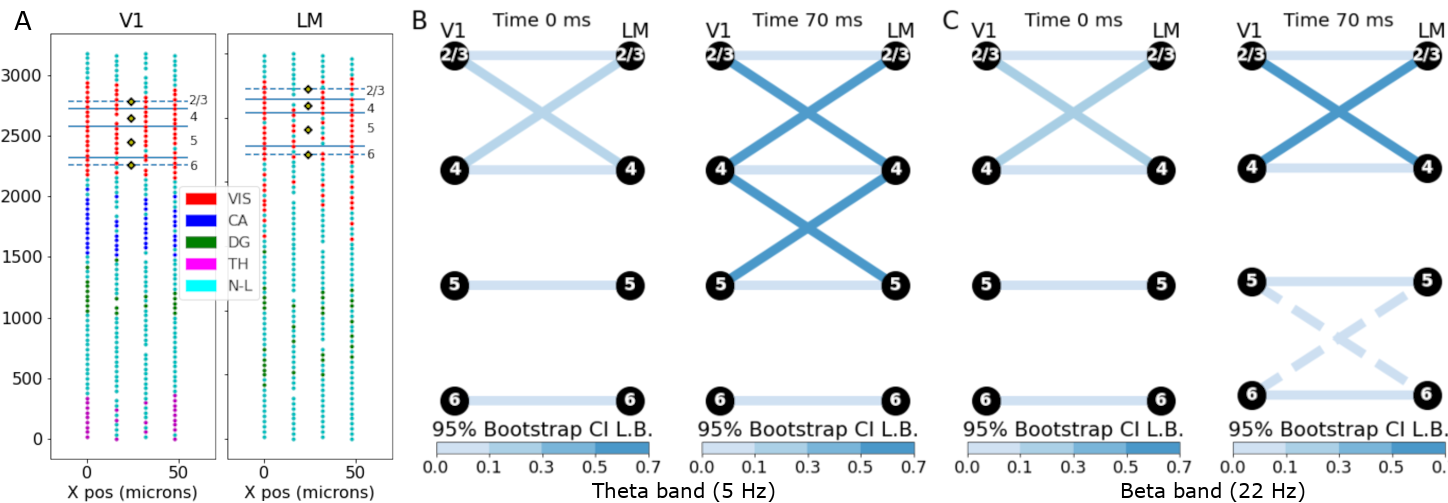}
    \caption{{\bf Phase coupling in Neuropixels data.} (A) Neuropixels probe LFP electrode locations (circles) for V1 (left) and LM (right), colored by putative region (VIS: visual cortex, CA: Cornu Ammonis, DG: dentate gyrus, TH: thalamus, N-L: no label). Putative cortical layer boundaries are overlaid on the red visual area electrodes with layer numbers indicated along the right side. Along the center of the probe are the locations we chose for CSD estimation (yellow diamonds), with one location centered in each cortical layer.
    (B) Torus graph phase coupling graphs based on theta oscillations at two time points relative to the stimulus. Edges shown for torus graph $p < 0.0001$. Edge color indicates the lower bound of a 95\% bootstrap confidence interval on the partial PLV value. It appears the strongest edges were between V1 and LM at similar cortical depths, with some evidence that edges were stronger at 70ms than at 0ms. (C) Similar to B, but for beta oscillations; dashed edges indicate edges with weaker evidence (torus graph $p < 0.001$). Similar to theta band, the connection patterns were mostly across similar layers but appeared slightly stronger at 70ms compared to 0ms.}
    \label{fig:neuropixels}
\end{figure}

The visual cortex data set was obtained from an experiment in which six Neuropixels probes were simultaneously inserted into the left hemisphere of a mouse brain~\cite{siegle2021survey}. The probes were targeted to the retinotopic centers of primary visual cortex (V1) and five higher-order visual areas (AM, PM, LM, AL, and RL), and extended down into portions of the hippocampus, thalamus, and midbrain. For each probe, the lower boundary of cortex was identified based on a decrease in the density of detected units approximately 800 µm below the brain surface. We analyzed the LFP data from 150 trials in which a 250 ms flash stimulus was presented to the right visual hemifield.
In this data set, we were interested in investigating relationships between V1 and higher-order visual regions~\cite{glickfeld2017higher}; as an example, we analyze phase connectivity between V1 and LM.
We first applied the GPCSD method (assuming a zero-mean Gaussian process) to the cortical electrodes from each of the probes to infer the latent CSDs corresponding to steady activity (with the average evoked response subtracted from the LFPs), then used torus graphs~\cite{torusgraphs} to determine phase connectivity.
We chose to predict the CSD at locations along the center of the probe that would correspond to the center of each putative cortical layer, depthwise. 
The Neuropixels probe recording locations, putative cortical visual layer boundaries, and CSD prediction locations are shown in Fig~\ref{fig:neuropixels}A. 

For phase coupling analysis, we selected two frequency bands of interest, theta band (centered at 5 Hz) and beta band (centered at 22 Hz). 
Similar to the torus graph analysis of steady auditory potentials and CSDs, the predicted CSDs were filtered using Butterworth bandpass filters with plus or minus 2 Hz width and the instantaneous phases were extracted for each trial using the Hilbert transform.
We chose two time points of interest, 0 ms (stimulus onset) and 70 ms (during the stimulus) to estimate torus graphs.
Figs~\ref{fig:neuropixels}B and~\ref{fig:neuropixels}C show the torus graph results, with edges shown for edgewise $p < 0.0001$ (solid) and $p < 0.001$ (dashed). 
To quantify edge strength, we computed the partial PLV statistic, which depends on the torus graph parameters and, like PLV, falls in the range $[0, 1]$.
Furthermore, we recomputed the partial PLV statistic based on fitted torus graphs across 1,000 bootstrap resamplings of the trials to measure uncertainty in the partial PLV statistic.
The edges in Figs~\ref{fig:neuropixels}B and~\ref{fig:neuropixels}C are colored by the lower bound of the 95\% bootstrap confidence interval. 
In both frequency bands, the primary connectivity between V1 and LM was between layers at similar depths (e.g., layer 5 to layer 5).
For both the theta band (Fig~\ref{fig:neuropixels}B) and beta band (Fig~\ref{fig:neuropixels}C), the connections were similar across time points but appeared stronger after stimulus compared to stimulus onset~\cite{sherman2016neural}.

\section*{Discussion}
We developed the GPCSD framework to improve not only spatial localization of currents, but also assessments of cross-population coupling on a trial-by-trial basis. LFPs result from the mixing of signals propagated from many current sources. Estimation with GPCSDs deconvolves these current source signals, rendering spatiotemporal processes that are more sensitive to analysis than the original LFPs.
This is apparent in Figs~\ref{fig:CSDspectra}B and~\ref{fig:CSDTGgraph}A, where analysis of CSD, but not LFP, reveals fluctuations in alpha power across cortical layers and, then, layer-to-layer phase coupling across populations recorded by the two probes. In addition, Fig~\ref{fig:1dshiftgraph} displays highly statistically significant correlations, across particular sources, in their
trial-specific timing of transient evoked responses, with consistent lead-lag relationships. It is not possible to see this kind of trial-by-trial evoked-response connectivity using LFPs directly. 

In several places we have compared and contrasted the GPCSD framework with that of kCSD.  In statistical parlance the kernel approach, in this setting, amounts to regularized nonparametric regression, which is well-established and reasonable. We find
GPCSD to be simple and powerful, but we do not mean to imply that it is uniquely compelling. 
Rather, in the settings we have described, it can capture important features of the data and, because it is both intuitive and flexible, others could develop it further, or tailor it to their own specific contexts, in fairly obvious ways.
Computation time would be greatly reduced, and occasional difficulties in optimization may be mostly eliminated, if the Gaussian process parameters and spatial extent radius $R$ were pre-specified. This could be based on other, related data, or it could be based on improved methods for initialization. The benefit comes from computing the integrals in Eq~(\ref{eq:marglik}) only once (as in kCSD) rather than iteratively, when predicting steady state activity.

We have also modeled spontaneous activity as the sum of two processes labeled ``slow'' and ``fast.'' This came from our observation that a sum of two such processes could often reproduce the spectra of spontaneous LFPs. The decomposition into slow and fast is, of course, supposed to be easy to comprehend, and it turns out to be adequate for many purposes. It does provide a nice interpretation, but, like so many other data analytic procedures, the two-process decomposition contains some arbitrariness and we do not mean to reify it. The extent to which it corresponds to reality is an empirical question that will rarely be settled by data of this type.
We have found cases where local maxima of the likelihood function do not nicely separate slow and fast components. In our implementation, we sought to address this problem with the standard solution of using multiple random initial values and retaining only the solution with the highest likelihood. Better methods of initialization could, again, help. In addition, future work could examine specific biophysical models that generate LFPs, thereby establishing ground truth, and the ability of GPCSD to correctly identify biologically meaningful characteristics of current sources. Finally, future implementations might also enforce charge conservation (a balance of current sources and sinks) or allow spatially-specific temporal covariance structure to replace the separable spatiotemporal kernel we have used here. We hope to investigate these ideas.

Eqs~(\ref{eq:phi})~and~(\ref{eq:tildephi}) take the form of a standard ``signal plus noise'' statistical model: CSD is the unobserved signal to which noise is added in producing observed LFPs. Eq~(\ref{eq:csdgenmodel}) models the CSD in a familiar additive form, while Eq~(\ref{eq:lfpgenmodel}) takes advantage of the linearity of the forward operator. This linearity, together with the Gaussian process assumptions, make GPCSD both flexible and tractable. Our analysis of simulated and real data was intended to display the kinds of results the framework can enable. It is possible that alternative formulations, with variations on this theme, could do even better. We hope the research reported here will lead to better CSD evaluations and new findings involving cross-population functional connectivity.

\section*{Methods and materials}
\label{Methods}
\subsection*{Ethics statement}
The treatment of the monkeys was in accordance with the guidelines set by the U.S. Department of Health and Human Services (NIH) for the care and use of laboratory animals, and all methods were approved by the Institutional Animal Care and Use Committee at the University of Pittsburgh.

\subsection*{Default GPCSD priors and optimization}
In this section, we briefly describe how the default GPCSD priors are set up; unless specified otherwise below, the default priors and model settings were used for all applications to real and simulated data. 

The default prior for the one-dimensional forward model parameter $R$ is inverse Gamma with 1\% and 99\% quantiles set to the minimum distance between electrodes and half the maximum distance between electrodes, respectively. The parameter is also by default bounded to fall between half the minimum distance between electrodes and 0.8 times the maximum distance between electrodes, as we found extremely small or large $R$ could cause numerical issues and are unlikely to be correspond to reasonable descriptions of the data.

The default spatial covariance is squared exponential (SE) with lengthscale prior set as inverse Gamma with 1\% and 99\% quantiles set to 1.2 times the minimum difference between electrode locations and 0.8 the maximum difference between electrode locations, respectively. During optimization, the lengthscale is bounded to fall between half the minimum distance between electrodes and the maximal distance between electrodes, to avoid lengthscales outside the range that the data can meaningfully inform. The spatial integrals needed for the spatial covariance functions are by default calculated using Gauss-Legendre quadrature with bounds equal to the edges of the electrode spatial location(s) and a default of 100 integration points in each dimension.

The default temporal covariance is a sum of SE and exponential terms. The SE lengthscale is given an inverse Gamma prior with 1\% and 99\% quantiles set to 1.2 times the minimum difference between time points and 0.8 times the maximum difference between time points, respectively; the variable is bounded during optimization similarly to the spatial lengthscale (to prevent values smaller than the data spacing or larger than the entire span of time points). The exponential lengthscale is by default given the same prior as the squared exponential, but more informative priors can be used to encourage each process toward a particular lengthscale. The default prior for the variances is half-Normal with standard deviation equal to 2, which is intended to be highly uninformative as long as the LFP data is scaled to approximately unit variance before estimation. Similarly, the LFP white noise variance is given a half-Normal prior with standard deviation 0.5.

Optimization of hyperparameters is accomplished using the L-BFGS-B algorithm as implemented in scipy~\cite{2020SciPy-NMeth} with default optimizer settings as suggested in~\cite{basak2021numerical}. To obtain the Jacobian of the objective function, we used automatic differentiation via the HIPS/autograd Python package \url{https://github.com/HIPS/autograd}. For robust optimization, the code by default repeats each optimization multiple times with different starting values for the hyperparameters randomly sampled from their prior distributions, and the set of hyperparameters resulting in the largest marginal log likelihood is retained.
Numerical results shown in the paper were produced on a Macintosh personal computer with 2.4 GHz 8-core Intel Core i9 processors and 32 GB RAM.

\subsection*{kCSD settings}
In our comparisons of GPCSD to 1D kCSD, we used the kCSD-python package \url{https://github.com/Neuroinflab/kCSD-python/releases/tag/v2.0}~\cite{potworowski2012kernel, pettersen2006current}. We used the cross-validation capability to tune the hyperparameters, and when possible, provided either the ground truth $R$ cylinder radius value or used the $R$ estimated by GPCSD (as kCSD does not provide an avenue for estimating this value). For the 1D simulation studies, we used the default number (1000) and spacing of Gaussian basis functions, allowing the basis function width to vary from 100 to 800 (15 total values) and the noise/regularization lambda value was varied between $10^{-15}$ and 1, with 25 total values (logarithmically spaced). In each application, we verified that the selected parameters were not on the boundary of the parameter grid.

\subsection*{Simulation details}
The simple CSD template was set up on a spatial grid spanning a spatial dimension with minimum value 0 microns and maximum value 2400 microns and a temporal dimension with 50 integer-valued time points.
The CSD template was made up of two positive-valued unit magnitude Gaussian-shaped bumps (means: 200 and 1600 in spatial dimension, 25 and 30 in temporal dimension; SDs: 150 in spatial dimension, 3 and 4 in temporal dimension) and two negative-valued unit magnitude Gaussian-shaped bumps (means: 800 and 2200 in spatial dimension, 25 and 30 in temporal dimension; SDs: 150 in spatial dimension, 3 and 4 in temporal dimension). 
The CSD template, evaluated at 50 time points and 2400 spatial locations, was passed through the one-dimensional forward model with parameter $R = 150$ (using the trapezoid rule to compute the integral) to generate a noiseless LFP observed at 24 spatial locations between 0 and 2400 microns across the 50 time points. 
The white noise variance was $\sigma^2 = 7 \times 10^{-5}$.
To fit GPCSD to the simple template, the default priors were used with 10 random starting values drawn from the prior (where the model with the best log likelihood was retained).
We found that in both the noiseless and white noise cases, the GPCSD models selected similar cylinder radius values ($R = 166, R = 160$), and similar spatial covariance lengthscales (219 and 220), temporal lengthscales (SE: 4.4 and 4.5; exponential: 17.5 and 17.5), and temporal variances (SE: $1.6 \times 10^{-6}$ and $1.8 \times 10^{-6}$; exponential near zero in both cases). In the noiseless case, the white noise variance was near zero, while in the noisy case it was estimated as $6.7 \times 10^{-5}$, which was very close to the true value. The kCSD selected basis widths were 550 in both the noiseless and noisy cases; the corresponding lambdas were 0.0002 and 0.004, respectively.

For the multi-trial simulation, kCSD used cross-validation in the same manner as the simple template simulation (with $R$ set to the ground truth value), and because the kCSD method applies independently to each time step and does not explicitly make use of multiple trials as independent realizations, trials were concatenated before performing cross-validation.
For computational reasons, only the first five trials were used for kCSD estimation, as computation time increased greatly if all trials were used. The selected basis width was 164.3 and the selected lambda was $3 \times 10^{-15}$.

GPCSD used MAP estimation with the default prior settings and ten restarts to the optimization. The estimated parameter values did not exactly match the generating parameters, so that GPCSD exhibited predictive performance indistinguishable from kCSD when the true parameters were used but was slightly worse when the estimated parameters were used.
The null hypothesis that the mean MSEs across trials were the same for each pair of methods (GPCSD and tCSD, GPCSD and kCSD) were tested using paired $t$-tests.

\subsection*{Auditory LFP data details}
The auditory LFP recordings consist of LFPs from two 24-electrode linear probes (V-Probes from Plexon) inserted in primary auditory cortex of a macaque monkey.
The probes were arranged parallel to the iso-frequency bands in primary auditory cortex (A1), and had similar tonal response fields with preferred frequencies close to 1000 Hz. The first probe (which we call the lateral probe) was located centrally in A1, while the second probe (which we call the medial probe) was located more medially and closer to the boundary of A1 with the medio-lateral belt. The medial probe had lower response threshold, shorter MUA latencies, and overall stronger current sinks and sources than the lateral probe.
The spacing between electrodes on each probe was 100 microns so that the probe spanned 2,300 microns.
The sampling rate of the LFPs was 1000 Hz and there were at least 2000 trials (stimulus presentations) per session.

Stimuli consisted of short tones at different frequencies and with different latencies between tones across the session.
The tones were 80 dB and lasted 55ms, with 11 different frequencies spaced linearly in $\log_2$ space (starting at 257 Hz and increasing by 0.32740 octaves each step).
The times since the last tone onset (the \textit{inter-stimulus intervals}, or ISIs) ranged between 0.2 and 18.0 seconds and followed a box-car distribution in $\log_2$ space.
The particular neural populations being recorded in each session will be tuned to respond to particular preferred frequencies, and trials with longer ISIs tend to elicit larger responses~\cite{pereira2014effects,teichert2016contextual}, though the mechanisms underlying this well-known variation in response amplitude with ISI are currently still being debated.

The auditory paradigm was designed to test whether neural responses in auditory cortex are modulated by whether or not the time or identity of a tone can be anticipated.
To that aim, the paradigm alternated between blocks of trials in which the identity and ISIs between tones was fixed for long sequences, and blocks of trials in which they were not.
Detailed analyses that are not presented here have revealed subtle effects of predictability on neural responses in auditory cortex.
However, because the effect of predictability is small, this analysis includes all trials from the predictable as well as the unpredictable blocks.
We focused on a single session in a single animal in which simultaneous recordings were made using two probes, spaced 3mm apart and with similar frequency tuning, so that relationships between the probes could be studied.
Trials containing large artifacts were not used in the analysis; in particular, trials in which the difference between the voltage at channels 1 and 24 exceeded 500 microvolts in absolute value in either probe were removed, resulting in 2,509 trials for the two-probe simultaneous recording.
The auditory data analyzed in this paper is available~\cite{teichert_tobias_2021_5137888}.
While in principle the conductivity scalar $\varsigma$ of Eq~(\ref{eq:poisson}) should be measured experimentally, in this work, we focused on recovering the spatiotemporal pattern of the CSDs, so we used $\varsigma = 1$ and treated all CSD estimates as having arbitrary units. 
Multi-unit activity was estimated as the envelope of power in the frequency range between 500 and 3000 Hz. To that aim, the data was first band-pass filtered in this frequency range, rectified, and then filtered with a 100 Hz low-pass.

\subsection*{Auditory analysis details}
The spatial covariance hyperparameters used the default GPCSD priors, but the extent of the integral was extended outside the electrode array by 200 microns in each direction to account for the fact that the LFPs may contain contributions from outside the electrodes; we found this to perform better in practice on real data though it was not necessary in simulations.
The temporal SE lengthscale prior had quantiles at 30ms and 100ms and the temporal exponential lengthscale prior had quantiles at 1ms and 20ms, to encourage the SE component to extract slow-timescale variation and the exponential component to extract fast-timescale variation.
The forward model parameter $R$ and the signal and noise variances had the default GPCSD priors.

Based on the periodograms, we selected a frequency range with a dominant oscillation, then filtered each channel and trial using a fourth-order Butterworth bandpass filter centered at the frequency of interest plus or minus 2 Hz; the filter was applied forward and backward to prevent phase distortion, then the Hilbert transform of the filtered signals was used to extract instantaneous phases at each time point.
Based on timecourses of Phase Locking Value~\cite{lachaux1999measuring} aggregated across all possible between- and within-probe electrode pairs, we selected a time point of interest to further investigate phase coupling.
Phase coupling at this time point was assessed using torus graphs~\cite{torusgraphs} on all between- and within-probe electrode pairs.
We first performed model selection to determine whether a torus graph submodel would be preferable to the full torus graph model, then tested whether there appeared to be any across-probe connections (null hypothesis of no connections) using a stringent alpha level of 0.0001.
We followed this test with edgewise tests for all within- and between- connections (edgewise alpha level of 0.01 with Bonferroni correction for all edges tested). 

To segment the CSD evoked response, local maxima were detected in the absolute value of the CSD evoked response, then an image segmentation watershed algorithm~\cite{beucher1993morphological} was applied to find clusters around each maxima; maxima that were above or below the range of electrode locations were excluded.
Space-time points not belonging to any component were kept the same for each trial and had no amplitude or shift variation.

To discourage physiologically implausible shift values, per-component shift values were restricted to no more than 20 ms in magnitude.
Observed correlations between the point estimates of the shifts of different components, computed across trials, were transformed using Fisher's $z$ transform, then $p$-values were obtained using Normal quantiles; significant correlations were determined using an alpha level of 0.001 with Bonferroni correction for all pairs tested.
Lagged relationships between the seven earliest evoked components in each probe were established using paired $t$-tests using an alpha level of 0.001 with Bonferroni correction for all pairs tested.

\subsection*{Neuropixels LFP data details}
The data were collected using Neuropixels probes, following the procedures described in~\cite{siegle2021survey}. The data is part of the Allen Brain Observatory Neuropixels dataset (copyright 2019 Allen Institute for Brain Science, available from \url{https://portal.brain-map.org/explore/circuits/visual-coding-neuropixels}). We analyzed one experiment in which six Neuropixels probes were simultaneously inserted through visual cortex, hippocampus, thalamus, and midbrain. On each probe, LFP data was recorded from up to 374 electrode locations in a checkerboard layout spanning two spatial dimensions: four columns spaced 16 microns apart, with 20 micron spacing between rows. LFP data was acquired at 2500 Hz after applying a 1000 Hz low-pass filter. Boundaries between regions were manually identified based on decreases in unit density as well as physiological signatures (such as elevated theta-band activity in the hippocampus). LFP electrodes without region labels were not included in the analysis. The visual stimulus consisted of 150 trials of simple black or white full-field flashes on a mean-luminance gray background.
Spike trains and LFP data for this mouse (subject ID: 730760270; session ID: 755434585) can be accessed on Zenodo~\cite{klein_natalie_2021_5150708} or as part of the Allen Brain Observatory AWS Public Data Set (\url{https://registry.opendata.aws/allen-brain-observatory/}).

\section*{Acknowledgements}
JHS thanks the Allen Institute founder, Paul G. Allen, for his vision, encouragement, and support.

\section*{Supporting information}

\subsection*{S1 Text. Modeling and computational details and additional simulation results.}
\label{S1Text}

\subsubsection*{One-dimensional \textit{a priori} physical model details}
For completeness, we give the derivation of the forward model, which was originally proposed in \cite{pettersen2006current}.
Substituting the \textit{a priori} cylinder model into the three-dimensional forward model yields
\begin{align}
    \phi(x,y,z) = - \frac{1}{4\pi\varsigma} \int_a^b {\int \int}_{x^2 + y^2 \le R}  \frac{g(z')}{\sqrt{(x-x')^2+(y-y')^2+(z-z')^2}} \, dx' \, dy' \, dz'.
\end{align}
We assume $x$ and $y$ are inside the cylinder (as typically, we assume we observe $\phi$ at the center of the cylinder). 
Changing to polar coordinates, we define $r^2 = (x-x')^2+(y-y')^2$ as the variable radius inside the cylinder and use the substitution $dx' \, dy' \, dz' = r \, d\theta \, dr \,  dz'$ to obtain
\begin{align}
    \phi(x,y,z) &= - \frac{1}{4\pi\varsigma} \int_a^b \int_0^R \int_0^{2\pi}  \frac{r g(z')}{\sqrt{(z-z')^2+r^2}} \, d\theta \, dr  \, dz' \\
    &= - \frac{1}{2\varsigma} \int_a^b g(z') \int_0^R \frac{r }{\sqrt{(z-z')^2+r^2}} \, dr \, dz' \\
    &= - \frac{1}{2\varsigma} \int_a^b g(z') \left[ \sqrt{(z-z')^2+R^2} - \sqrt{(z-z')^2}\right] \, dz'.
\end{align}
Notice that after integration, this is no longer a function of $x$ or $y$, so we can simply write
\begin{align}
    \phi(z) = - \frac{1}{2\varsigma} \int_a^b g(z') \left[ \sqrt{(z-z')^2+R^2} - \sqrt{(z-z')^2}\right] \, dz'.
\end{align}
To better understand how $R$ affects the $\phi$, we can factor out $R$:
\begin{align}
     \phi(z) = - \frac{R}{2\varsigma} \int_a^b g(z') \underbrace{\left[ \sqrt{\left(\frac{r}{R}\right)^2+1} - \sqrt{\left(\frac{r}{R}\right)^2}\right]}_{a(z, z'; R)} \, dz' 
\end{align}
where $r = z - z'$ and $a(z, z';R)$ is a weight function with a maximum value of 1 when $r = 0$.
Typically, we are interested only in the relative magnitude of the CSD in space and time so the scalar $\frac{R}{2\varsigma}$ can be ignored.

\subsubsection*{Two-dimensional \textit{a priori} physical model details}
As mentioned in the text, certain choices of two-dimensional \textit{a priori} physical models may lead to singularities in the forward model.
To avoid the singularity, we assume that there is some region of zero CSD surrounding the probe, which is parameterized by $\tau$. 
As shown by \cite{lkeski2011inverse}, substituting the \textit{a priori} model into the three-dimensional forward model yields
\begin{align*}
    \phi(x,y,z) &= - \frac{1}{4\pi\varsigma}  \int_{a_z}^{b_z} \int_{a_y}^{b_y} \int_{\tau \le x \le R+\tau}  \frac{g(y', z')}{\sqrt{(x-x')^2+(y-y')^2+(z-z')^2}} \, dx' \, dy' \, dz' \\
    &= - \frac{1}{4\pi\varsigma}  \int_{a_z}^{b_z} \int_{a_y}^{b_y} \int_{\tau}^{R+\tau}  \frac{g(y', z')}{\sqrt{(x-x')^2+m^2}} \, dx' \, dy' \, dz' \\
    &= - \frac{1}{4\pi\varsigma} \int_{a_z}^{b_z} \int_{a_y}^{b_y} g(y', z') \int_{\tau}^{R+\tau} \frac{1}{\sqrt{(x-x')^2+m^2}} \, dx' \, dy' \, dz'
\end{align*}
where $m = \sqrt{(y-y')^2+(z-z')^2}$.
Since we are interested in modeling the LFP at the face of the probe $(x = 0)$, let $x = 0$ and integrate over $x'$:
\begin{align*}
    \phi(0,y,z) &= - \frac{1}{4\pi\varsigma}  \int_{a_z}^{b_z} \int_{a_y}^{b_y} g(y', z') \int_{\tau}^{R+\tau} \frac{1}{\sqrt{(x')^2+m^2}} \, dw \, dy' \, dz' \\
    &= - \frac{1}{4\pi\varsigma}  \int_{a_z}^{b_z} \int_{a_y}^{b_y} g(y', z') \left[ \log(R + \tau + \sqrt{(R+\tau)^2 + (y-y')^2+(z-z')^2} \right. \\
    & \hspace{4em} - \left. \log(\tau + \sqrt{\tau^2 + (y-y')^2+(z-z')^2}\right] \,  dy' \, dz'.
\end{align*}
We will write the LFP as $\phi(y,z)$ where implicitly $x = 0$ when using the forward model.
As in the one-dimensional case, $\frac{1}{4\pi\varsigma}$ may be dropped if one is only interested in the relative variation of the CSD across space and time (and in many cases, $\varsigma$ may not be known). 

\subsubsection*{Computational details for Gaussian process}
Let $\mbf{K}^s \in \R^{M \times M}$ be the LFP spatial covariance evaluated at the locations of the observed LFPs and $\mbf{K}^t \in \R^{T \times T}$ be the temporal covariance evaluated at the observed time points; they are functions of $\bs{\theta}$ (including the forward model parameter $R$ which is part of the LFP spatial covariance function through the forward operator; the computation of $\mbf{K}^s$ is discussed in the next section).
Let $\trial{\bs{\tilde{\phi}}} \in \R^{M \times T}$ represent the matrix of observed LFPs on trial $n$ (with $N$ total trials).
The traditional expression for the log marginal likelihood (excluding terms that don't depend on $\bs{\theta}$) takes the form
\begin{align*}
    \log \mathcal{L}(\bs{\theta}) = - \frac{1}{2} \sum_{n=1}^N \log \left( \left| \mbf{K}^s \otimes \mbf{K}^t + \sigma^2 \mbf{I} \right| \right) + \text{vec}\left(\trial{\bs{\tilde{\phi}}}\right)^T \left[\mbf{K}^s \otimes \mbf{K}^t + \sigma^2 \mbf{I} \right]^{-1}\text{vec}\left(\trial{\bs{\tilde{\phi}}}\right).
\end{align*}
However, this form relies on inversion of an $MT \times MT$ matrix which is clearly problematic for typical $M$ and $T$ observed in real data, so we instead leverage the special structure present in the matrix.
In particular, we use the eigendecomposition of the covariance matrices, $\mbf{K}^s = \mbf{Q}_s \mbf{\Lambda}_s \mbf{Q}_s^T$ and $\mbf{K}^t = \mbf{Q}_t \mbf{\Lambda}_t \mbf{Q}_t^T$, where $\mbf{\Lambda}_s$ and $\mbf{\Lambda}_t$ are diagonal matrices.
This is useful because of the following identity:
\begin{align*}
    \left[\mbf{K}^s \otimes \mbf{K}^t + \sigma^2 \mbf{I} \right]^{-1} &= 
    \left[\mbf{Q}_s \mbf{\Lambda}_s \mbf{Q}_s^T \otimes \mbf{Q}_t \mbf{\Lambda}_t \mbf{Q}_t^T + \sigma^2 \mbf{I} \right]^{-1} \\
    &= \left[(\mbf{Q}_s \otimes \mbf{Q}_t) (\mbf{\Lambda}_s \otimes \mbf{\Lambda}_t) (\mbf{Q}_s^T \otimes  \mbf{Q}_t^T) + \sigma^2 \mbf{I} \right]^{-1} \\
    &= \left[(\mbf{Q}_s \otimes \mbf{Q}_t) (\mbf{\Lambda}_s \otimes \mbf{\Lambda}_t) (\mbf{Q}_s^T \otimes  \mbf{Q}_t^T) + \sigma^2 (\mbf{Q}_s \otimes \mbf{Q}_t) (\mbf{Q}_s \otimes \mbf{Q}_t)^{-1} \right]^{-1} \\
    &= \left[(\mbf{Q}_s \otimes \mbf{Q}_t) (\mbf{\Lambda}_s \otimes \mbf{\Lambda}_t) (\mbf{Q}_s^T \otimes  \mbf{Q}_t^T) + (\mbf{Q}_s \otimes \mbf{Q}_t) (\sigma^2 \mbf{I}) (\mbf{Q}_s^T \otimes \mbf{Q}_t^T) \right]^{-1} \\
    &= \left[(\mbf{Q}_s \otimes \mbf{Q}_t) (\mbf{\Lambda}_s \otimes \mbf{\Lambda}_t + \sigma^2 \mbf{I}) (\mbf{Q}_s^T \otimes  \mbf{Q}_t^T)   \right]^{-1} \\
    &= (\mbf{Q}_s \otimes \mbf{Q}_t) \left[ \mbf{\Lambda}_s \otimes \mbf{\Lambda}_t + \sigma^2 \mbf{I} \right]^{-1} (\mbf{Q}_s^T \otimes  \mbf{Q}_t^T)
\end{align*}
which relies on properties of the Kronecker product and the orthonormality of $(\mbf{Q}_s \otimes \mbf{Q}_t)$.
This implies that after eigendecomposition, the inversion of the matrix reduces to inversion of a diagonal matrix.

Let $\mbf{D} = \mbf{\Lambda}_s \otimes \mbf{\Lambda}_t + \sigma^2 \mbf{I}$ be a diagonal matrix, and let $\mbf{q}$ be an $MT$-vector with elements $1/D_{ii}$.
Note that a low-rank Gaussian process could be implemented by using truncated eigendecompositions \cite{solin2014hilbert}.
Using the eigendecomposition and properties of Kronecker products (as shown in more detail in \cite{saatcci2012scalable}), the log marginal likelihood may be rewritten:
\begin{align*}
    \log \mathcal{L}(\bs{\theta}) = -\frac{N}{2} \sum_{i=1}^{MT} \log(D_{ii}) - \frac{1}{2} \sum_{n=1}^N \sum_{i=1}^{MT} \left[\text{vec}\left({\mbf{Q}_s^T \trial{\bs{\tilde{\phi}}} \mbf{Q}_t}\right) \circ \text{vec}\left({\mbf{Q}_s^T \trial{\bs{\tilde{\phi}}} \mbf{Q}_t}\right) \circ \mbf{q} \right]_i
\end{align*}
where the Hadamard product $\circ$ indicates elementwise multiplication and $\mbf{Q}_t$, $\mbf{Q}_s$, $\mbf{D}$, and $\mbf{q}$ depend on $\bs{\theta}$.
This form is faster and more stable to compute as it avoids direct inversion of an $MT \times MT$ matrix which has computational complexity $O(M^3T^3)$.
Because eigendecomposition is of complexity $O(M^3)$ and $O(T^3)$ for the spatial and temporal covariance matrices, respectively, the computational complexity of one likelihood function evaluation is instead $O(M^3 + T^3 + MT)$ for both eigendecompositions and the inversion of a diagonal matrix of size $MT \times MT$.

Given fixed $\bs{\theta}$, the log marginal likelihood may also be optimized over mean function parameters $\bs{\gamma}$; here we show the likelihood assuming a shared mean function across trials, though per-trial mean parameters could also be used (and if trials were assumed independent, this would result in a separate log marginal likelihood for each trial).
Let $\bs{\mu} \in \R^{M \times T}$ be the mean function evaluated at the observed LFP spatial and temporal points (where this function depends on $\bs{\gamma}$).
We first calculate the inverse covariance matrix as
\begin{align*}
    \bs{\Sigma}^{-1} = \left( \mbf{Q}_s \otimes \mbf{Q}_t \right) \text{diag}(\mbf{q}) \left( \mbf{Q}_s \otimes \mbf{Q}_t \right)^T
\end{align*}
and calculate the mean of the LFPs across trials as $\bar{\mbf{y}} = \frac{1}{N_r} \sum_{r=1}^{N_r} \trial{\bs{\tilde{\phi}}}$.
Then we use the following (rescaled) log marginal likelihood:
\begin{align*}
    \log \mathcal{L}(\bs{\gamma}) = \text{vec}(\bs{\mu})^T \bs{\Sigma}^{-1} \text{vec}(\bar{\mbf{y}}) - \frac{1}{2}  \text{vec}(\bs{\mu})^T \bs{\Sigma}^{-1} \text{vec}(\bs{\mu}).
\end{align*}

\subsubsection*{Numerical integration in computing covariance matrices}
To compute the spatial LFP covariance matrix, the forward model must be applied to the CSD covariance function and evaluated at the observed LFP spatial locations.
We will assume that the integral is approximated using a standard numerical integration scheme of the form
\begin{align*}
    \int_a^b f(u) \, du \approx \sum_i w_i f(u_i) 
\end{align*}
where $w_i$ are weights that may depend on a quadrature scheme or the distance between the $u_i$ points.
In the case of a covariance function for the LFP, we apply the forward model integral equation to both inputs of the covariance function, so that the function evaluated at a single pair of inputs $(x, x')$ takes the form
\begin{equation}
    \int_a^b \int_a^b b(x - u) b(x' - v) k(u, v) \, du \, dv \approx \sum_i \sum_j w_i^u w_j^v b(x - u_i) b(x' - v_j) k(u_i, v_j)
\label{eq:numint}
\end{equation}
where $k$ is the CSD covariance function and $b$ are the forward model weights.
Assume we want to evaluate the LFP covariance matrix at all pairs of locations from two vectors $\mbf{x} = [x_1, ..., x_m]$ and $\mbf{x'} = [x_1, ..., x_n]$, and assume we have vectors $\mbf{u} = [u_1, ..., u_c]$ and $\mbf{v} = [v_1, ..., v_d]$ spanning the ranges of the integrals.
Define the following matrices:
\begin{align*}
    \mbf{A} &= \begin{bmatrix}
        w_1^u b(x_1 - u_1) & \cdots & w_c^u b(x_1 - u_c) \\
        w_1^u b(x_2 - u_1) & \cdots & w_c^u b(x_2 - u_c) \\
        \vdots & \vdots & \vdots \\
        w_1^u b(x_m - u_1) & \cdots & w_c^u b(x_m - u_c)
    \end{bmatrix} \in \mathbb{R}^{m \times c}, \\
    \mbf{B} &= \begin{bmatrix}
        w_1^v b(x'_1 - v_1) & \cdots & w_1^v b(x'_n - v_1) \\
        w_2^v b(x'_1 - v_2) & \cdots & w_2^v b(x'_n - v_2) \\
        \vdots & \vdots & \vdots \\
        w_d^v b(x'_1 - v_d) & \cdots & w_d^v b(x'_n - v_d)
    \end{bmatrix} \in \mathbb{R}^{d \times n}, \\
    \mbf{K} &= \begin{bmatrix}
        k(u_1, v_1) & \cdots & k(u_1, v_d) \\
        k(u_2, v_1) & \cdots & k(u_2, v_d) \\
        \vdots & \vdots & \vdots \\
        k(u_c, v_1) & \cdots & k(u_c, v_d)
    \end{bmatrix} \in \mathbb{R}^{c \times d}.
\end{align*}
Then the LFP spatial covariance may be computed as $\mbf{A}\mbf{K}\mbf{B}$.
Similarly, the spatial cross-covariance between the LFP and the CSD may be computed as $\mbf{A}\mbf{K}$.
We found that using simple integration rules (midpoint or trapezoid rule) worked well given large enough $c$ and $d$.
Notice that the multiplication $\mbf{A}\mbf{K}\mbf{B}$ must be done prior to the eigendecomposition, or else the orthonormality need to establish the key identity for computing the matrix inverse is not preserved. 

The scheme described above also applies directly to the two-dimensional case. 
To evaluate a single element of the covariance matrix at spatial locations $(x_1, x_2)$ and $(x'_1, x'_2)$, where single subscripts now represent dimension indexing, we evaluate the integral
\begin{align*}
    \int_{a_2}^{b_2} \int_{a_2}^{b_2} \int_{a_1}^{b_1} \int_{a_1}^{b_1} b(x_1 - u_1, x_2 - u_2) b(x'_1 - v_1, x'_2 - v_2) k( u_1, u_2, v_1, v_2) \, du_1 \, dv_1 \, du_2 \, dv_2. 
\end{align*}
Now assuming $\mbf{u}$ and $\mbf{v}$ are two-dimensional grids indexed as $u_{k,i}$ and $v_{k,j}$ where $k$ is the dimension index in $\{1, 2\}$ and $i, j$ are the element indices, the integral can be approximated as
\begin{align*}
    \sum_i \sum_j w^u_i w^v_j b(x_1 - u_{1, i}, x_2 - u_{2, i}) b(x'_1 - v_{1, j}, x'_2 - v_{2, j}) k( u_{1, i}, u_{2, i}, v_{1, j}, v_{2, j})
\end{align*}
which is again a double sum and can be written in the form $\mbf{A}\mbf{K}\mbf{B}$.

In evaluating the likelihood and making predictions with the Gaussian process, the spatial covariance matrix must be calculated using the numerical integration schemes discussed in this section.
Given $n_s$ spatial points in the integration grid, computing the spatial covariance matrix requires $n_s^2$ calculations.
While $n_s$ is chosen by the user, a small value leads to inaccurate numerical integration. In addition, while moderate $n_s$ may be reasonable for one-dimensional GPCSD, two-dimensional GPCSD generally requires larger $n_s$ because it is the total number of points in a two-dimensional grid, so is typically roughly quadratic in the number of points required for one-dimensional integration.
If the forward model parameter $R$ and the Gaussian process parameters were known, the spatial covariance matrix could be computed once with a cost of $O(n_s^2)$ and the computational complexity of following likelihood evaluations would be $O(M^3 + T^3 + MT)$ where $M$ is the number of observed spatial points and $T$ is the number of observed time points.
However, the numerical integration needed to compute the spatial covariance matrix must be repeated each time the spatial covariance or forward model parameters change, leading to $O(n_s^2 + M^3 + T^3 + MT)$ for each likelihood evaluation as those parameters are varied.

\subsubsection*{Additional simulation results}
In addition to simulating from a GPCSD model then fitting a GPCSD model with the same model form, we also investigated simulating from a GPCSD model and fitting a mis-specified GPCSD model.
We looked at two cases; for each case, we again generated 50 trials each with 60 time points and 24 spatial locations (similar to the auditory LFP probe).

First, we fit a GPCSD model with only a squared exponential temporal covariance plus LFP white noise to data generated from a squared exponential (lengthscale 20, variance 1.5),  plus Mat\'ern (lengthscale 2, variance 0.2), plus LFP white noise. 
The average MSE across 50 trials was 0.01, which is orders of magnitude higher than we found using a correctly specified model in the main text. 
Visually, the estimated CSD captured the overall slow-timescale trend, but failed to capture fast non-white-noise fluctuations. 
These results suggest that models that are not flexible enough may fail to fit the data well.
While lack of fit (in the LFP space) can be used to diagnose this issue, and models with varying numbers of components can be compared by their negative log likelihood values, it may be difficult to pin down the exact number of components that best explains a particular data set.
That is, diagnostics can suggest a more flexible model is needed, but in the spirit of model parsimony, decisively selecting a minimal number of components may require a subjective judgement.

Second, we fit a GPCSD model with two components (squared exponential and Mat\'ern) to data generated using three components (fast squared exponential with lengthscale 10 and variance 0.5, slow squared exponential with lengthscale 100 and variance 0.1, and Mat\'ern with lengthscale 2 and variance 0.2).
In this case, the average MSE across trials was $7.4 \times 10^{-5}$, similar to the error with a correctly specified model in the main text.
This case study demonstrates that if the fitted model form is flexible enough, it can achieve a good fit to the data even if the underlying true generating process contained more components (which may be difficult to disentangle based on real data if, for instance, their temporal spectral properties overlap).

\end{document}